\DeclareSymbolFont{AMSb}{U}{msb}{m}{n}
\documentclass[reqno,centertags,10pt,a4paper,noamsfonts]{amsart}
\pdfoutput=1
\usepackage[svgnames]{xcolor}
\usepackage{graphicx}
\usepackage[english]{babel}
\usepackage[babel=true,expansion=alltext,protrusion=alltext-nott,final]{microtype}
\usepackage[margin={3cm},marginparwidth={2.5cm}]{geometry}
\usepackage[utf8]{inputenc}
\usepackage{textcomp}
\usepackage[sb]{libertine}
\usepackage[libertine,bigdelims,vvarbb]{newtxmath}
\useosf
\usepackage[varqu,varl]{inconsolata}
\usepackage[supsfam=LibertinusSerif-Sup,%
       supscaled=1.2,%
       raised=-.13em]{superiors}
\usepackage{mathtools}
\usepackage[T1]{fontenc}
\usepackage{textcomp}
\usepackage{amsthm}

\usepackage[cal=cm]{mathalpha}
\usepackage[inline]{enumitem}
\numberwithin{equation}{section}
\usepackage[normalem]{ulem}

\usepackage{float}  
\usepackage{caption}
\usepackage{subcaption}
\usepackage{xspace}         
\usepackage[scientific-notation=true]{siunitx}      
\usepackage{pstricks}
\usepackage{tikz}
\usetikzlibrary{positioning} 
\usetikzlibrary{calc}
\usepackage{pgfplots}
\pgfplotsset{width=10cm,compat=1.9}
\usepackage{bm}
\usepackage{sidecap}
\usepackage{graphicx,color}

\usepackage{amsmath}
\DeclareFontFamily{U}{mathx}{}
\DeclareFontShape{U}{mathx}{m}{n}{<-> mathx10}{}
\DeclareSymbolFont{mathx}{U}{mathx}{m}{n}
\DeclareMathAccent{\widehat}{0}{mathx}{"70}
\DeclareMathAccent{\widecheck}{0}{mathx}{"71}

\providecommand{\mr}[1]{\href{http://www.ams.org/mathscinet-getitem?mr=#1}{MR~#1}}
\providecommand{\zbl}[1]{\href{https://zbmath.org/?q=an:#1}{Zbl~#1}}

\providecommand{\doi}[2]{\href{https://doi.org/#1}{#2}}

\newcommand{\C}{\mathcal{C}}

\newcommand{\ii}{\imath}

\definecolor{light_gray}{gray}{0.75}
\definecolor{lighter_gray}{gray}{0.5}
\colorlet{light_blue}{blue!20}
\definecolor{dark_green}{rgb}{0.0, 0.6, 0.0}
\definecolor{royal_blue}{rgb}{0.0, 0.22, 0.66}
\definecolor{salmon}{rgb}{1.0, 0.55, 0.41}
\definecolor{gold}{rgb}{0.8, 0.63, 0.21}
\definecolor{navy_blue}{rgb}{0.0, 0.0, 0.5}
\definecolor{crimson}{rgb}{0.79, 0.0, 0.09}
\definecolor{amethyst}{rgb}{0.6, 0.4, 0.8}
\definecolor{alizarin}{rgb}{0.82, 0.1, 0.26}
\definecolor{amaranth}{rgb}{0.9, 0.17, 0.31}
\definecolor{azure}{rgb}{0.0, 0.5, 1.0}
\definecolor{canaryyellow}{rgb}{0.82, 0.41, 0.12}
\definecolor{carrotorange}{rgb}{0.8, 0.33, 0.0}
\definecolor{cadmiumgreen}{rgb}{0.0, 0.42, 0.24}
\definecolor{copper}{rgb}{0.72, 0.45, 0.2}
\definecolor{aqua}{rgb}{0.5, 1.0, 0.83}
\definecolor{awesome}{rgb}{1.0, 0.13, 0.32}
\definecolor{candyapplered}{rgb}{1.0, 0.03, 0.0}
\definecolor{caribbeangreen}{rgb}{0.0, 0.8, 0.6}
\definecolor{indigo}{rgb}{0.0, 0.25, 0.42}

\DeclareMathOperator{\weaklystar}{\rightharpoonup\kern-2.2ex ^* \, \,}

\def\XXint#1#2#3{{\setbox0=\hbox{$#1{#2#3}{\int}$ }
\vcenter{\hbox{$#2#3$ }}\kern-.6\wd0}}

\newcommand{\R}{\mathbb R}
\newcommand{\N}{\mathbb N}
\newcommand{\Z}{\mathbb Z}
\renewcommand{\C}{\mathbb C}

\newcommand\norm[1]{\lVert #1 \rVert}
\newcommand\bignorm[1]{\big\lVert #1 \big\rVert}

\newcommand\inner[1]{\langle #1 \rangle}

\newcommand\scpr{\boldsymbol{\cdot}}

\newcommand{\ra}{\rightarrow}

\newcommand{\mL}{\mathrm{L}}

\renewcommand{\phi}{\varphi}

\newcommand{\mH}{\mathrm{H}}
\newcommand{\mW}{\mathrm{W}}

\newcommand{\T}{\mathbb{T}}

\newcommand{\ee}{\mathrm{e}}

\theoremstyle{plain}
\newtheorem{theorem}{Theorem}[section]
\newtheorem{proposition}[theorem]{Proposition}

\newtheorem{lemma}[theorem]{Lemma}

\newtheorem*{theorem*}{Theorem}

\theoremstyle{definition}
\newtheorem{definition}[theorem]{Definition}
\newtheorem{remark}[theorem]{Remark}
\newtheorem*{remark*}{Remark}

\usepackage[pagebackref=false]{hyperref}
\hypersetup{
	linktoc=all,
	bookmarksdepth=3,
	colorlinks=true,
	linkcolor=Green,citecolor=Orange,urlcolor=DarkBlue,
	pdftitle={A rigorous formulation of DFT},
	pdfauthor={Thiago Carvalho Corso}, 
	pdfsubject={},
	pdfkeywords={}} 
\begin{document}
\numberwithin{table}{section}
\title[A rigorous formulation of DFT]{A rigorous formulation of Density Functional Theory for spinless fermions in one dimension}

\author[T.~Carvalho~Corso]{Thiago Carvalho Corso}
\address[T.~Carvalho Corso]{Institute of Applied Analysis and Numerical Simulation, University of Stuttgart, Pfaffenwaldring 57, 70569 Stuttgart, Germany}
\email{thiago.carvalho-corso@mathematik.uni-stuttgart.de}


\thanks{\emph{Funding information}:  DFG -- Project-ID 442047500 -- SFB 1481.  \\[1ex]
\textcopyright 2025 by the authors. Faithful reproduction of this article, in its entirety, by any means is permitted for noncommercial purposes.}
\begin{abstract}
In this paper, we present a completely rigorous formulation of Kohn-Sham density functional theory for spinless fermions living in one dimensional space. More precisely, we consider Schr\"odinger operators of the form
\begin{align*}
    H_N(v,w) = -\Delta + \sum_{i\neq j}^N w(x_i,x_j) + \sum_{j=1}^N v(x_i) \quad \mbox{acting on $\bigwedge^N \mL^2([0,1])$,}
\end{align*}
where the external and interaction potentials $v$ and $w$ belong to a suitable class of distributions. In this setting, we obtain a complete characterization of the set of pure-state $v$-representable densities on the interval. Then, we prove a Hohenberg-Kohn theorem that applies to the class of distributional potentials studied here. Lastly, we establish the differentiability of the exchange-correlation functional and therefore the existence of a unique exchange-correlation potential. We then combine these results to provide a rigorous formulation of the Kohn-Sham scheme. In particular, these results show that the Kohn-Sham scheme is rigorously exact in this setting. 
\end{abstract}
\setcounter{tocdepth}{1}
\maketitle
\tableofcontents
\setcounter{secnumdepth}{2}

\section{Introduction}

In this paper we present a mathematically rigorous derivation of Kohn-Sham Density Functional Theory (DFT) as an exact ground-state theory.

\subsection{Motivation} Density functional theory (DFT) has established itself as a cornerstone of modern quantum chemistry, solid-state physics, and material science. By offering a computationally efficient alternative to wavefunction-based methods, DFT has become the most widely used method for large scale electronic structure calculations, see, e.g., \cite{KBP96,Bur12, Jon15, ED11, CF23} and references therein.

At the heart of DFT lies the Kohn-Sham scheme, which seeks to reproduce the ground-state (single-particle) density of an interacting system of electrons via a fictitious (or effective) system of non-interacting electrons. In the original formulation of Kohn and Sham \cite{KS65}, this is achieved by introducing the so-called exchange-correlation functional, reformulating the ground-state variational problem over the set of Slater determinants, and computing the first order optimality conditions (Euler-Lagrange equations).  These optimality conditions are the celebrated Kohn-Sham equations and can be written as a (nonlinear) eigenvalue equation for a non-interacting Hamiltonian, usually called the Kohn-Sham system. 

Although initially proposed as an approximate scheme, Kohn-Sham DFT is often termed a \emph{formally exact} theory. This widespread claim rests on the assumption that, if the exact derivative of the exchange-correlation functional -- the so-called exchange-correlation potential-- could be explicitly evaluated, then the resulting Kohn-Sham equations would lead to the exact ground-state density of any interacting system of interest. However, from a mathematical perspective, there are several potential pitfalls with this assumption:
\begin{enumerate}[label=(\arabic*)]
    \item (Existence) First, it is not clear whether, for a given interacting system, there exists a non-interacting (Kohn-Sham) system whose eigenfunction exactly reproduces the ground-state density of the interacting system. Typically, one expects this eigenfunction to be the ground-state of the Kohn-Sham system. This is called the \emph{Aufbau} principle and is not justified either.
    \item (Uniqueness) Second, if a Kohn-Sham system exists, it is not clear whether it is unique. Put differently, are there two non-interacting Hamiltonians whose ground-state (or excited state) single-particle densities are the same?
    \item (Regularity) Third, it is not clear whether the exchange-correlation functional is differentiable at all and the exchange-correlation potential well-defined. In particular, if a minimizer of the Kohn-Sham energy exists, in which sense does it satisfies the Euler-Lagrange equations?
\end{enumerate}

The aforementioned questions are not new and their importance is well recognized in the literature \cite{THS+22,WAR+23,PTC+23}. The first question is known as the $v$-representability problem and is paramount to a mathematically rigorous formulation of KS-DFT. More precisely, the $v$-representability problem consists in characterizing the set of all possible ground-state densities of Schr\"odinger operators of the form
\begin{align}
    H_N(v,w) = -\Delta + \sum_{i\neq j}^N w(x_i,x_j) + \sum_{j=1}^N v(x_j) \quad \mbox{acting on $\bigwedge^N \mL^2(\Omega)$,} \label{eq:Hamiltonian0}
\end{align}
for a fixed interaction operator $w$, and a class of external potentials $v$. To the best of the author's knowledge, this question is completely open in the case of three dimensional continuous systems, i.e., $\Omega \subset \R^3$. Nevertheless, in simplified settings such as finite and infinite lattice systems \cite{CCR85,PL21}, the $v$-representability question is well-understood. Moreover, in the case of continuous one-dimensional systems, some notable progress has been made \cite{AS88,CS91,CS93}. In particular, in a recent breakthrough paper by Sutter el al \cite{SPR+24}, the authors provided the first sufficient criteria for a function to be \emph{ensemble} $v$-representable, i.e., a convex combination of ground-state densities of $H_N(v,w)$, in the one-dimensional torus $\Omega = \T = \R /\Z$. This was done by extending the class of admissible external potentials to include certain distributions in dual Sobolev spaces. However, their work does not provide \emph{necessary} conditions for ensemble $v$-representability and do not address the \emph{pure}-state $v$-representability problem, which is in fact necessary in the original formulation of the Kohn-Sham method. 

The second question is related to the so-called Hohenberg-Kohn theorem \cite{HK64}. This problem has also received considerable attention in the literature \cite{Lie83,Lam18,Gar18,Gar19,LBP20}. It is currently known that, under suitable integrability assumptions on the class of admissible external potentials, the Kohn-Sham Hamiltonian associated to a given interacting system, if existing, is unique. However, these results do not apply to distributional potentials, e.g., the class of potentials studied in \cite{SPR+24}. Therefore, it is not clear whether, by extending the class of potentials to gain ensemble $v$-representability, one compromises the uniqueness of the Kohn-Sham system. 

The differentiability question has also been studied before \cite{Lam07,KET+14}. However, these works focus on the differentiability of the convex Lieb functional (or regularizations thereof), which can be seen as a relaxation (or convexification) of the celebrated Levy-Lieb constrained-search functional \cite{Lev79,Lie83}. While these functionals are certainly related to the exchange-correlation functional introduced by Kohn and Sham, neither of these results seems to address the differentiability of the latter, which is crucial for a rigorous understanding of the Kohn-Sham scheme.

We also remark that the aforementioned questions are not only relevant for the (forward) Kohn-Sham scheme but also for the inverse problem \cite{SW21}, whose ultimate goal is to control the ground-state density by tuning the external potential. Therefore, all three questions are of significant scientific interest. Nevertheless, despite their relevance, a satisfactory solution, even in the case of continuous one-dimensional systems, is still missing. It is therefore our main goal in this paper to fill in this gap.

\subsection{Main contributions}

In this paper we show that all of the aforementioned issues can be rigorously addressed in the setting of spinless fermions living in one-dimensional space, i.e., $\Omega = (0,1)$ in~\eqref{eq:Hamiltonian0}. More precisely, the main contributions of this paper can be summarized as follows:
\begin{enumerate}
    \item \label{it:v-rep} We provide a complete characterization of the set of \emph{pure-state} $v$-representable densities  in a bounded interval. More precisely, for a fixed interaction $w$, we provide necessary and sufficient conditions for a function to be the ground-state density of an operator of the form $H_N(v,w)$ under Neumann boundary conditions (BCs). In particular, we show that the set of non-interacting and $w$-interacting $v$-representable densities are the \emph{same} for any $w$ in a large class of distributional potentials. In addition, we provide a similar characterization in the case of periodic and anti-periodic BCs under an additional constraint on the number of particles.
    \item \label{it:HK} We prove a Hohenberg-Kohn theorem for distributional potentials, i.e., we show that, for a fixed interaction potential $w$, the external potential $v$ is uniquely determined by the ground-state density of $H_N(v,w)$. Combined with the previous result, this theorem shows that distributional potentials are not only sufficient but also necessary to represent a reasonable set of densities.
    \item \label{it:KS} We prove that the exchange-correlation functional introduced by Kohn and Sham is Gateaux differentiable, and consequently, the exchange-correlation potential exists and is well-defined. Moreover, we show that the Kohn-Sham kinetic energy is also differentiable. We then show that  the Kohn-Sham scheme can be rigorously formulated and the Aufbau principle holds.
\end{enumerate}

These results demonstrate that, in the one-dimensional setting, the ground-state density of any system of interacting fermions can be exactly reproduced within the Kohn-Sham framework. In particular, to the best of the author's knowledge, this provides the first rigorous proof of the widespread claim that Kohn-Sham DFT is an exact ground-state theory for continuous electronic systems.

\section{Results}
    In this section, we state our main results precisely. We then outline the key steps in the proofs and how these steps are organized throughout the paper.
\subsection{Notation} \label{sec:notation} We start with some notation. Throughout this paper, we let $I=(0,1)$ be the open unit interval and set $I_N \coloneqq (0,1)^N$ for any $N\in \N$. 

We denote by $\mH^1(I)$ the Sobolev space of functions $f\in\mL^2(I)$ with weak derivative $\partial_x f \in \mL^2(I)$. Moreover, for $1\leq p \leq \infty$ and $N\in \N$, we denote by $\mW^{1,p}(I_N)$ the  Sobolev spaces of functions in $\mL^p(I_N)$ with weak gradient in $\mL^p(I_N)$, and by $\mW^{-1,q}(I_N)$, where $1/q+1/p = 1$, the dual space of $\mW^{1,p}(I_N)$. In addition, we denote by $\mH^{1/2}(\partial I_N)$ the standard $1/2$-Sobolev (or Besov) space along the boundary $\partial I_N$.

We also denote by $\mH^1_{+1}(I_N)$ and $\mH^1_{-1}(I_N)$ the Sobolev spaces of periodic and anti-periodic functions, respectively. More precisely, $\Psi \in \mH^1_{\pm 1}(I_N)$ if and only if $\Psi \in \mH^1(I_N)$ and
\begin{align*}
    (\gamma\Psi)(x_1,...,x_{j-1},0,x_{j},...,x_{N-1}) = \pm (\gamma \Psi)(x_1,...,x_{j-1},1,x_j,...,x_{N-1}),
\end{align*}
for almost every $(x_1,...,x_{N-1}) \in I_{N-1}$ and every $1\leq j \leq N$, where $\gamma : \mH^1(I_N) \rightarrow \mH^{1/2}(\partial I_N)$ denotes the standard Dirichlet trace operator.

We define $\mathcal{V}$ as the following space of generalized external potentials:
\begin{align*}
\mathcal{V} \coloneqq \{ v \in \mH^{-1}(I) : v(\phi) \in \R \quad\mbox{for any real-valued $\phi\in \mH^1(I)$}\}.
\end{align*}
Similarly, we define $\mathcal{W}$ as the following space of generalized (pairwise) interaction potentials:
\begin{align}
    \mathcal{W} \coloneqq \{w \in \mW^{-1,q}(I_2): q>2\quad \mbox{and}\quad w(\phi) \in \R \quad \mbox{for any real-valued $\phi \in \mW^{1,p}(I_2)$}\}.
\end{align} 

For $N\in \N$, we denote by $\mathcal{H}_N$ the usual space of spinless electronic wave-functions, i.e., the antisymmetric $N$-fold tensor product
\begin{align*}
    \mathcal{H}_N \coloneqq \bigwedge^N \mL^2(I).
\end{align*}
For $v\in \mathcal{V}$ and $w\in \mathcal{W}$, we denote by $H_N(v,w)$ the $N$-particle Hamiltonian
\begin{align}
    H_N(v,w) = -\Delta + \sum_{i\neq j}^N w(x_i,x_j) + \sum_{j=1}^Nv(x_i) \quad \mbox{acting on $\mathcal{H}_N$.} \label{eq:Hamiltonian}
\end{align}
More precisely, we shall consider four different self-adjoint realizations of $H_N(v,w)$. To properly introduce these realizations, let us denote by $\mathfrak{a}_{v,w}$ the sesquilinear form
\begin{align}
    \mathfrak{a}_{v,w}(\Psi,\Phi) = \int_{I_N} \overline{\nabla \Psi(x_1,...,x_N)} \scpr \nabla \Phi(x_1,...,x_N) \mathrm{d}x_1...\mathrm{d} x_N + v\left(\rho_{\Psi,\Phi}\right) + w(\rho^{(2)}_{\Psi,\Phi}), \label{eq:quadratic form0}
\end{align}
where $\rho_{\Psi,\Phi}$ is the overlapping single-particle density 
\begin{align}
    \rho_{\Psi,\Phi}(x) \coloneqq N \int_{I_{N-1}} \overline{\Psi(x,x_2,...,x_N)} \Phi(x,x_2,...,x_N) \mathrm{d} x_2...\mathrm{d} x_N , \label{eq:density def}
\end{align}
and $\rho^{(2)}_{\Psi,\Phi}$ is the overlapping pair density
\begin{align}
    \rho^{(2)}_{\Psi,\Phi} \coloneqq N(N-1) \int_{I_{N-2}}\overline{\Psi(x,y,x_3,...,x_N)} \Phi(x,y,x_3,...,x_N) \mathrm{d}x_2...\mathrm{d}x_N. \label{eq:pair density def}
\end{align}
The fact that the pairing $v(\rho_\Psi)$ and $w(\rho^{(2)}_\Psi)$ is well-defined for any $\Psi \in \mathcal{H}_N \cap \mH^1(I_N)$ will be clarified in Section~\ref{sec:background}. Then we denote respectively by $H_N(v,w)$, $H_N^{+}(v,w)$, and $H_N^{-}(v,w)$ the unique self-adjoint operators associated to the sesquilinear form $\mathfrak{a}_{v,w}$ with the following form domains:
\begin{enumerate}[label=(\roman*)]
    \item(Neumann) $\quad \quad \mathcal{Q}_N \coloneqq \mH^1(I_N) \cap \mathcal{H}_N.$
    \item (Periodic) $\quad\quad  \,\,\,\, \mathcal{Q}^+_N \coloneqq \mH^1_{+1}(I_N)\cap \mathcal{H}_N$.
    \item (Anti-periodic) $\,\,\, \mathcal{Q}^-_N \coloneqq \mH^1_{-1}(I_N)\cap \mathcal{H}_N$.
\end{enumerate}
For further details on the construction of these self-adjoint operators, we refer to Section~\ref{sec:self-adjoint}.

\begin{remark}[Generalized Neumann boundary conditions]
Strictly speaking, functions in the domain of the Neumann realization $H_N(v,w)$ do not necessarily have a vanishing outward normal derivative along the boundary. The reason is that, since we allow for distributional potentials that could be supported on the boundary, the domain of $H_N(v,w)$ may correspond to functions satisfying Robin boundary conditions. For instance, this is the case if $v =\delta_0$ where $\delta_0$ is the Dirac measure at $x=0$, see e.g. \cite[Section 2]{BRS18}. 
\end{remark}

\begin{remark}[Periodic boundary conditions via the Torus] For periodic boundary conditions, we can equivalently identify the form domain with the set of $\mH^1$ functions in the $N$-dimensional Torus $\T^N = \R^N/\Z^N$. This is the setting considered in the previous works \cite{SPR+24,Cor25a}.
\end{remark}

\subsection{Main results~(\ref{it:v-rep}) - Characterization of $v$-representability}

In this section, we address the pure-state $\mathcal{V}$-representability problem.

Our first result is a complete characterization of the set of \emph{pure-state} $\mathcal{V}$-representable densities on the interval under Neumann boundary conditions. To the best of the author's knowledge, this is the first complete solution to the pure-state $\mathcal{V}$-representability problem for an infinite-dimensional and continuous system. 
\begin{theorem}[Characterization of pure-state $\mathcal{V}$-representability - Neumann BCs]\label{thm:v-rep} Let $w\in \mathcal{W}$ be fixed and $N\in \N$. Then the set of all possible ground-state\footnote{Unless otherwise stated, we always assume a ground-state wave-function to be normalized, i.e., $\norm{\Psi}_{\mL^2} =1$.} densities of the Neumann realization $H_N(v,w)$ for $v\in \mathcal{V}$, i.e., the set
    \begin{align}
        \mathcal{D}_N(w) \coloneqq \{ \rho_{\Psi} : \Psi \mbox{is a ground-state of $H_N(v,w)$ for some $v\in \mathcal{V}$}\}, \label{eq:pure-state local def}
    \end{align}
    is given by
    \begin{align}
        \mathcal{D}_N(w) = \left\{\rho \in \mH^1(I): \int_I \rho(x) \mathrm{d}x = N\quad \mbox{and}\quad \rho(x) >0 \quad \mbox{for any $x\in [0,1]$}\right\}. \label{eq:v-rep Neumann}
    \end{align}
    In particular, $\mathcal{D}_N(w) = \mathcal{D}_N$ is independent of the interaction potential $w\in\mathcal{W}$. 
    \end{theorem}

In the case of periodic or anti-periodic boundary conditions, we can also give a complete characterization of the set of pure-state $\mathcal{V}$-representable densities, but under an additional constraint on the number of particles.

\begin{theorem}[Characterization of pure-state $\mathcal{V}$-representability - non-local BCs] \label{thm:v-rep non-local} Let $w\in \mathcal{W}$ and denote by $\mathcal{D}_N^{\pm}(w)$ the sef of all possible ground-state densities of $H_N^\pm(v,w)$ for $v\in \mathcal{V}$, i.e.,
\begin{align}
    \mathcal{D}^{\pm}_N(w) \coloneqq \{ \rho_\Psi : \Psi \mbox{ is a ground-state of $H_N^\pm(v,w)$ for some $v\in \mathcal{V}$}\}. \label{eq:pure-state non-local def}
\end{align}
Then for any $k\in \N$, we have
\begin{align}
    \mathcal{D}^+_{2k-1}(w) = \mathcal{D}^+_{2k-1} \quad \mbox{and}\quad \mathcal{D}^-_{2k}(w) = \mathcal{D}^+_{2k}, \label{eq:v-rep non-local densities}
\end{align}
where 
\begin{align}
    \mathcal{D}_{N}^{+} \coloneqq \left\{ \rho \in \mH^1(I) : \int_I \rho(x) \mathrm{d} x = N, \quad \rho(0) = \rho(1), \quad \mbox{and}\quad \rho(x) > 0 \quad \mbox{for any $x\in [0,1]$}\right\}. \label{eq:v-rep Per}
\end{align}

\end{theorem}

\begin{remark}[Anti-periodic wave-functions have periodic density] Note that any anti-periodic wave-function has a periodic density, which justifies the $+$ sign on the right hand side of both equations in~\eqref{eq:v-rep non-local densities}.
\end{remark}

At this point, the reader may wonder whether the condition on the number of particles is merely an artifact of the proof. In other words, one may naturally conjecture that 
\begin{align}
    \mathcal{D}^\pm_N(w) = \mathcal{D}^+_N \quad \mbox{for any number of particles $N\in \N$.} \label{eq:conjecture}
\end{align}
Although we cannot fully address this question here, the simple example of a single free particle with anti-periodic boundary conditions shows that this conjecture is false for $N=1$. More precisely, by considering the one-dimensional Laplacian with anti-periodic boundary conditions, one can prove the following.
\begin{proposition}[Counter example for $N=1$] \label{prop:v-rep counterexample} Let $N=1$, then we have
\begin{align*}
    \mathcal{D}^-_1 \neq \mathcal{D}^+_1,
\end{align*}
where $\mathcal{D}^{\pm}_1$ are the sets defined in~\eqref{eq:pure-state non-local def}\footnote{Since there is no distinction between interacting and non-interacting $\mathcal{V}$-representability in the case of a single-particle, we simply write $\mathcal{D}^\pm_1$ instead of $\mathcal{D}^\pm_1(0)$ for the sets introduced in~\eqref{eq:pure-state non-local def}}.
\end{proposition}
Nevertheless, it was shown in \cite{Cor25a} that for \emph{non-interacting} periodic systems, the set of ensemble $\mathcal{V}$-representable densities is precisely $\mathcal{D}^+_N$, i.e., the set
\begin{align*}
    \mathcal{D}_{N,\rm ens}^{+}(0) := \left\{ \sum_{j} \lambda_j \rho_{\Psi_j} : 0 
    \leq \lambda_j \leq 1 \,\,\, \sum_{j} \lambda_j = 1, \,\,\, \mbox{$\{\Psi_j\}$ ground-states of $H_N^+(v,0)$ for some $v\in \mathcal{V}$}\right\}
\end{align*}
is equal to $\mathcal{D}_N^+$ for any $N \in \N$. In other words, the conjecture in~\eqref{eq:conjecture} is true for non-interacting ensemble $\mathcal{V}$-representability in the case of periodic systems. Here we improve this result to pure-state $\mathcal{V}$-representability. In fact, we show that~\eqref{eq:conjecture} holds for the set of non-interacting $\mathcal{V}$-representable densities with any number of particles $N\geq 2$ and both periodic and anti-periodic BCs. 
\begin{theorem}[Characterization of non-interacting pure-state $\mathcal{V}$-representability - non-local BCs]\label{thm:v-rep non-interacting} Let $\mathcal{D}^\pm_N(w)$ be the sets defined in~\eqref{eq:pure-state non-local def}. Then we have
\begin{align}
    \mathcal{D}_N^{\pm}(0) = \mathcal{D}_N^+ \quad \mbox{for any $N \geq 2$.}
\end{align}
\end{theorem}

\begin{remark}[Dirichlet BCs]
The case of Dirichlet BCs, i.e., the self-adjoint realization with form domain $\mathcal{Q}_N^0 \coloneqq \mH^1_0(I_N) \cap \mathcal{H}_N$ is rather subtle and we are not able to address it here. We shall comment more on this point later in Section~\ref{sec:conclusion}.
\end{remark}

\subsection{Main results~(\ref{it:HK}) - Hohenberg-Kohn for distributional potentials} We now turn to the question of whether the external potential $v$ can be uniquely reconstructed from the ground-state density. In the affirmative case, such result is known as the Hohenberg-Kohn (HK) theorem in the DFT literature. Here, we prove the following version of the HK theorem.
\begin{theorem}[Hohenberg-Kohn with distributional potentials] \label{thm:HK} Let $w\in \mathcal{W}$, $N\in \N$ and $v,v'\in \mathcal{V}$, and suppose that $H_N(v,w)$ and $H_N(v',w)$ have the same ground-state density $\rho \in \mathcal{D}_N$. Then $v-v'$ is constant.
\end{theorem}

In the case of periodic and anti-periodic BCs, the potential $v\in \mathcal{V}$ cannot be uniquely recovered from the ground-state density for the following reason. As $\mH^1_{+ 1}(I)$ is a proper closed subspace of $\mH^1(I)$, by the Hahn-Banach theorem\footnote{As $\mH^1_+(I)$ is a subspace of $\mH^1(I)$ with co-dimension one, we do not need to appeal to the Hahn-Banach theorem. In fact, any functional satisfying~\eqref{eq:equality periodic} is of the form $v' = \alpha(\delta_0 - \delta_1) +v$ for some $\alpha \in \R$, where $\delta_x$ denotes the Dirac's delta measure at $x\in [0,1]$}, there exists (infinitely) many functionals $v\neq v' \in \mathcal{V}$ such that
\begin{align}
    v(\rho) = v'(\rho) \quad \mbox{for any $\rho \in \mH^1_{+ 1}(I)$.} \label{eq:equality periodic}
\end{align}
In particular, for any such $v$ and $v'$ the operators $H_N^\pm(v,w)$ and $H_N^\pm(v',w)$ are the same and therefore have the same ground-state density. Of course, this lack of uniqueness only arises because we consider $\mathcal{V}$ as the real-valued distributions in the dual of $\mH^1(I)$ instead of the dual of $\mH^1_{+ 1}(I)$. Once we dismiss this artificial lack of uniqueness, the following version of the HK theorem holds.
\begin{theorem}[Hohenberg-Kohn theorem - non-local BCs] \label{thm:HK non-local} Let $w\in \mathcal{W}$, and $v,v' 
\in \mathcal{V}$ be such that the ground-state densities of $H_N^+(v,w)$ and $H_N^+(v',w)$ (respectively $H_N^-(v,w)$ and $H_N^-(v',w)$) are the same. In addition, suppose that $N$ is odd (respectively even). Then $(v-v')\rvert_{\mH^1_{+1}(I)}$ is constant, i.e., 
\begin{align*}
    v(\rho) = v'(\rho) + C \int_I \rho(x)\mathrm{d}x \quad \mbox{for any $\rho \in \mH^1_{+1}(I)$,}
\end{align*}
and some constant $C>0$ independent of $\rho$.
\end{theorem}

In view of Theorems~\ref{thm:HK} and~\ref{thm:HK non-local}, one may again be tempted to conjecture that the HK theorem holds for any number of particles under both periodic and anti-periodic boundary conditions. In the case of non-interacting periodic systems, this was indeed shown by the author in \cite{Cor25a}. Moreover, this result can be extended to the case of anti-periodic boundary conditions for any number of particles $N\geq 2$. However, somewhat surprisingly, the HK theorem does not hold for the case of a single-particle with anti-periodic boundary conditions. In this case the following simple counter example exists.
\begin{proposition}[Counter example to HK for $N=1$] \label{prop:counterexample} Let $v = \delta_{1/2}$, then the function $\rho(x) = 2\cos(\pi x)^2$, is a ground-state density of the self-adjoint realizations of $h^-(0) = - \Delta$ and $h^-(v) = - \Delta +v$ with form domain $\mH^1_{-1}(I)$.
\end{proposition}

\subsection{Main results~(\ref{it:KS}) - the Kohn-Sham scheme} We can now provide a rigorous formulation of the Kohn-Sham scheme. For simplicity, we state this result only for the Neumann case. The analogous results in the case of periodic and anti-periodic boundary conditions is highlighted later in Remark~\ref{rem:KS non-local}.

We begin by recalling some definitions. First, we define the Levy-Lieb constrained search functional as
\begin{align}
    F_{\rm LL}(\rho;w) \coloneqq \begin{dcases} \inf_{\Psi \in \mathcal{Q}_N \mapsto \rho} \{\norm{\nabla \Psi}_{\mL^2}^2 + w(\rho^{(2)}_{\Psi})\}, \quad &\mbox{for $\rho \in \mathcal{R}_N$,}\\
    +\infty, \quad &\mbox{otherwise,} \end{dcases} \label{eq:LL functional}
\end{align}
where $\mathcal{R}_N$ is the set of $N$-representable densities
\begin{align*}
    \mathcal{R}_N = \left\{ \rho: \sqrt{\rho} \in \mH^1(I), \quad \rho \geq 0, \quad \int_I \rho(x) \mathrm{d} x = N \right\}.
\end{align*}
Next, we define the Hartree functional $E_H: \mH^1(I) \rightarrow \C$ as
\begin{align*}
    E_H(\rho;w) \coloneqq w(\rho \otimes \rho), \quad \mbox{where}\quad (\rho \otimes \rho)(x,y) = \rho(x) \rho(y).
\end{align*}
Moreover, the Kohn-Sham kinetic energy functional $T_{\rm KS}: \mathcal{R}_N \rightarrow [0,\infty)$ can be defined as
\begin{align}
    T_{\rm KS}(\rho) = \min_{\substack{\Psi \in \mathcal{S}_N \\ \Psi \mapsto \rho}}  \int_{I_N} |\nabla \Psi(x_1,...,x_N)|^2 \mathrm{d} x_1 ... \mathrm{d} x_N, \label{eq:KS kinetic}
\end{align}
where $\mathcal{S}_N$ is the set of Slater determinants with finite kinetic energy, i.e., the set of all wave-functions of the form
\begin{align*}
   \Psi(x_1,...,x_N) = (\phi_1 \wedge ... \phi_N)(x_1,...,x_N) = \frac{1}{\sqrt{N!}} \sum_{\sigma \in \mathcal{P}_N} \mathrm{sgn}(\sigma) \phi_1(x_{\sigma(1)}) \phi_2(x_{\sigma(2)}) ... \phi_N(x_{\sigma(N)}), 
\end{align*}
for a collection of $\mL^2$-orthorgonal functions $\{\phi_j\}_{j=1}^N$ in $\mH^1(I)$. 

Using the above definitions, we can define the exchange-correlation functional $E_{\rm xc}: \mH^1(I;\R)\rightarrow \R \cup \{+\infty\}$ as
\begin{align}
    E_{\rm xc}(\rho;w) \coloneqq F_{\rm LL}(\rho;w) - T_{\rm KS}(\rho) - E_H(\rho;w). \label{eq:xc def}
\end{align}
The first result of this section is then the following.
\begin{theorem}[Differentiability of the exchange-correlation] \label{thm:xc} Let $E_{\rm xc}$ be the exchange-correlation functional defined in~\eqref{eq:xc def}. Then for any $w\in \mathcal{W}$, the map $\rho \mapsto E_{\rm xc}(\rho;w)$ is Gateaux differentiable at any point $\rho \in \mathcal{D}_N$, i.e., there exists a unique (up to an additive constant) potential $v_{\rm xc}(\rho) = \mathrm{d}_\rho E_{\rm xc} \in \mathcal{V}$ such that
\begin{align*}
    \lim_{\epsilon \ra 0^+} \frac{E_{\rm xc}(\rho+\epsilon \delta) - E_{\rm xc}(\rho)}{\epsilon} = v_{\rm xc}(\rho)(\delta),
\end{align*}
for any $\delta \in \mH^1(I;\R)$ with $\int_I \delta(x) \mathrm{d} x = 0$.
\end{theorem}

\begin{remark}[Tangent space of densities and differential of exchange-correlation] We shall see later that $\mathcal{D}_N$ is a relatively open subset of the affine space $N+\mathcal{X}_0$, where
\begin{align*}
    \mathcal{X}_0 = \left\{ \delta \in \mH^1(I;\R): \int_I \delta(x) \mathrm{d}x = 0\right\}.
\end{align*}
Hence, we can say that $\mathcal{X}_0$ is the "tangent space" to $\mathcal{D}_N$ at any $\rho \in \mathcal{D}_N$. In particular, we can naturally identify the "cotangent space" at any $\rho \in \mathcal{D}_N$ with the quotient space 
\begin{align*}
    \mathcal{V}/\{1\} \coloneqq \{[v] : v\sim v' \quad \mbox{if and only if} \quad v-v'=\mbox{constant}\}.
\end{align*}
Therefore, strictly speaking, $\mathrm{d}_\rho E_{\rm xc}$ is not a single potential but an equivalence class of potentials modulo additive constants.
\end{remark}

\begin{remark}[Hartree energy with pairwise interaction] Typically, the electron-electron interaction energy is given by integration of the pair density against $w(x-y)$ for some symmetric function  $w$ (i.e., $w(x) = w(-x)$). In this case, the Hartree distributional potential in~\eqref{eq:Hartree potential} is given by integration against the function
\begin{align*}
    v_H(\rho)(x) = \int_I w(x-y) \rho(y) \mathrm{d} y = (w\ast \rho)(x).
\end{align*}
This is the usual formula for the Hartree potential that appears throughout the DFT literature \cite{ED11, CF23}. Note that, for general $w\in \mathcal{W}$, one can still define the Hartree potential as in equation~\eqref{eq:Hartree potential} above.
\end{remark}
The last main result of this paper shows that the Kohn-Sham scheme is rigorously exact and the Aufbau principle holds. Precisely, we have 
\begin{theorem}[Exact Kohn-Sham DFT] \label{thm:KS-DFT} Let $v\in \mathcal{V}$, $w\in \mathcal{W}$ and $N\in \N$, and denote by $\rho(v;w)$ the (unique) ground-state density of $H_N(v,w)$. Then, there exists a unique (up to a global phase) minimizer of the Kohn-Sham energy
\begin{align*}
    \min_{\substack{\Psi \in \mathcal{S}_N \\ \norm{\Psi} =1 }}\left\{ \norm{\nabla \Psi}_{\mL^2}^2 + E_{\rm xc}(\rho_{\Psi}) + E_H(\rho_{\Psi}) + v(\rho_{\Psi})\right\}.
\end{align*}
Moreover, this minimizer is given by the Slater determinant of the $N$ lowest eigenfunctions of the Kohn-Sham single-particle Hamiltonian
\begin{align*}
   h(\rho(v;w)) = -\Delta + v_{\rm xc}(\rho(v;w)) + v_H(\rho(v;w)) + v,
\end{align*}
where $v_{\rm xc}(\rho) = \mathrm{d}_\rho E_{\rm xc} \in \mathcal{V}$ and $v_H(\rho) \in \mathcal{V}$ is the Hartree (distributional) potential
\begin{align}
   \delta \mapsto  v_H(\rho)(\delta) = w(\rho \otimes \delta) + w(\delta \otimes \rho). \label{eq:Hartree potential}
\end{align} 
\end{theorem}

\begin{remark}[Non-local BCs] \label{rem:KS non-local} The results in Theorem~\ref{thm:xc} and~\ref{thm:KS-DFT} also apply to the case of periodic and anti-periodic boundary conditions. To be precise, in these cases the spaces of densities $\mathcal{R}_N$ and potentials $\mathcal{V}$ have to be replaced by their periodic counterparts 
\begin{align*}
    \mathcal{R}_N^+ \coloneqq \mH^1_{+1}(I) \cap \mathcal{R}_N \quad \mbox{and}\quad \mathcal{V}_{+1} \coloneqq \mH^{-1}_{+1}(I;\R),
\end{align*}
and the minimization on the Kohn-Sham kinetic energy $T_{\rm KS}$ and constrained-search functional $F_{\rm LL}$ have to be restricted to periodic or anti-periodic wave-functions. With these modifications, and under the condition that the number of particles is odd in the periodic case and even in the anti-periodic one, the analogous statements of Theorems~\ref{thm:xc} and Theorems~\ref{thm:KS-DFT} hold. 
\end{remark}

\subsection{Key steps of the proofs}

The proof of Theorem~\ref{thm:v-rep} is divided into two steps. In the first step, we show that any density satisfying the conditions in~\eqref{eq:v-rep Neumann} are $\mathcal{V}$-representable. This step is a straightforward adaptation of the convex analysis argument in \cite{SPR+24}, used to prove sufficient conditions for ensemble $\mathcal{V}$-representability on the torus. The reason we obtain the stronger \emph{pure-state} $\mathcal{V}$-representability here is that the ground-state of $H_N(v,w)$ is non-degenerate. This non-degeneracy result was proved recently by the author in \cite{Cor25b}, and will be used several times throughout this paper. The second step of the proof consists in showing that any ground-state density of $H_N(v,w)$ satisfy the conditions in~\eqref{eq:v-rep Neumann}. The proof of the regularity condition $\rho \in \mH^1(I)$ is rather simple and standard, so the novel part is to show that the density is everywhere non-vanishing. For this, the key observation is that a vanishing point of the density implies the vanishing of the wave-function along a hyperplane crossing $I_N$. Therefore, we can combine Courant's nodal domain theorem \cite{Cou23} with the non-degeneracy theorem to show that the density does not vanish inside the interval. To show that the density does not vanish on the end points of the interval, we apply a weak unique continuation result along the boundary, cf. Theorem~\ref{thm:UCPboundary}.

The proof of Theorem~\ref{thm:v-rep non-local} follows similar steps, but with one significant change. In view of the non-local BCs, the opposite boundary faces of $\partial I_N$ are "glued" to each other, and therefore, we cannot split the domain with a single hyperplane. 
Consequently, we cannot apply Courant's theorem to show that the density is non-vanishing inside the interval $(0,1)$. 
Fortunately, in the case of periodic boundary conditions, every point in the torus is the same up to a translation. Therefore, we can translate any hyperplane parallel to the coordinate axis to the boundary and apply the weak unique continuation result to prove that the density is non-vanishing at any point $x\in [0,1]$. In the case of anti-periodic BCs, there is a similar but slightly more involved domain rearranging argument that allows us to do the same. Interestingly, this argument also allows us to slightly strengthen the weak unique continuation result proved in \cite[Theorem 6.1]{Cor25b} in the following way. 
\begin{theorem}[Improved weak unique continuation for periodic and anti-periodic BCs] \label{thm:UCPboundary2} Let $v\in \mathcal{V}$, $w\in \mathcal{W}$, and $\Psi$ be a ground-state of $H_N^+(v,w)$ for $N$ odd or $H_N^-(v,w)$ for $N$ even. Then $\Psi$ cannot vanish identically on any relatively open subset of a hyperplane parallel to one of the boundary faces of $\partial I_N$.
\end{theorem}


The proof of Theorem~\ref{thm:v-rep non-interacting} essentially follows from three observations. First, any density in $\mathcal{D}_N^+$ is periodic and anti-periodic ensemble $\mathcal{V}$-representable. This was shown in \cite{SPR+24} for periodic systems and follows from the proof of Theorem~\ref{thm:v-rep non-local} for anti-periodic ones. Second, the ground-state of the non-interacting operator $H_N^{\pm}(v,0)$ is at most two-fold degenerate. This follows from the fact that every eigenvalue of the single-particle operator $h^\pm(v) = -\Delta +v$ is at most double degenerate (see Theorem~\ref{thm:single-particle}), and can be used to show that any density in $\mathcal{D}_N^+$ is in fact non-interacting \emph{pure-state} $\mathcal{V}$-representable. Third, any ground-state density of the non-interacting system with $N\geq 2$ particles is pointwise larger than the ground-state density with $N =1$ and $N=2$ particles for periodic and anti-periodic BCs, respectively. In particular, the ground-state density for $N\geq 2$ is nowhere vanishing by Theorem~\ref{thm:v-rep non-local}.

The proof of Theorems~\ref{thm:HK} and~\ref{thm:HK non-local} also requires some new ideas. The standard approach used to prove previous versions of the Hohenberg-Kohn theorem \cite{Lie83,Gar18,Lam18, Gar19} is based on two steps: first, one uses a variational argument to conclude that the operators $H_N(v,w)$ and $H_N(v',w)$ have a mutual ground-state $\Psi$. This is the standard Hohenberg-Kohn argument \cite{HK64} and can be applied here as well. The second step consists in dividing the difference Schr\"odinger equation 
\begin{align}
    \left(H_N(v,w)  - H_N(v',w)\right) \Psi =\left( \lambda_1(v',w) - \lambda_1(v,w)\right)\Psi, \label{eq:dif SE}
\end{align}
where $\lambda_1(v,w)$ denotes the ground-state energy of $H_N(v,w)$, 
by the ground-state wave-function to infer that the potential is constant. In previous works, this division is possible in an almost everywhere pointwise sense because the potentials are multiplicative and the strong unique continuation principle (UCP) is known (or proven) to hold. While the UCP for the class of Schr\"odinger operators considered here may not hold for general weak solutions, it does hold for the ground-state wave-function by Theorem~\ref{thm:non-degenerate}. However, this is not enough to carry out the division argument. This can be seen, for instance, by considering a delta potential $v= \delta_x$ and a wave-function $\Psi$ that vanishes on the hyperplane $\{x\}\times I_{N-1}$. 

To overcome this issue, we proceed as follows. First, we show that~\eqref{eq:dif SE} is equivalent to $v-v'$ being an eigenfunction of an operator $K$ whose integral kernel is given by the pair density and the inverse of the single-particle density of $\Psi$. By exploiting the regularity of the pair density (Lemma~\ref{lem:regularity reduced densities}) and the fact that the density is nowhere vanishing, we can show that this operator $K$ is regularity improving. This allows us to prove that $v-v'$ is, in fact, a multiplicative potential in $\mL^1(I)$. Therefore, we are able to use the ground-state UCP in Theorem~\ref{thm:non-degenerate} to carry out the division step mentioned previously.

As a consequence of Theorems~\ref{thm:HK} and~\ref{thm:HK non-local}, one can also show that the ground-state energy is strictly monotone with respect to the external potential. This result is not directly related to DFT, but complements the strict monotonicity result with respect to enlarging the Dirichlet set, proved in \cite[Theorem 2.7]{Cor25b}. 
\begin{theorem}[Strict monotonicity of ground-state energy with respect to external potential] \label{thm:monotone} Let $w\in \mathcal{W}$ and $v,v'\in \mH^1(I)$ be such that $(v-v')\geq 0$ and $(v-v')\neq 0$, then the ground-state energies $\lambda_1(v,w)$ and $\lambda_1(v',w)$ satisfy
\begin{align*}
    \lambda_1(v,w) > \lambda_1(v'w).
\end{align*}
Moreover, if $(v-v')\rvert_{\mH^1_{+1}(I)} \neq 0$, the same result holds for periodic BCs when $N$ is odd and anti-periodic BCs when $N$ is even. 
\end{theorem}
Finally, the proof of Theorem~\ref{thm:xc} combines the non-degeneracy theorem and Theorem~\ref{thm:HK} with the observation made in \cite[Corollary 19]{SPR+24} that a Hohenberg-Kohn theorem is essentially equivalent to differentiability of the Lieb convex functional. More precisely, we use the Hohenberg-Kohn theorem established here to show that the constrained search functional has a unique subgradient at any $\rho \in \mathcal{D}_N$ and is therefore (Gateaux) differentiable. From this and the non-degeneracy result in Theorem~\ref{thm:non-degenerate}, we can then show that both the Kohn-Sham kinetic energy and the Hartree functional are differentiable at any point in $\mathcal{D}_N$. The proof of Theorem~\ref{thm:KS-DFT} follows by combining these results. 

\subsection{Outline of the paper}

In the next section we recall the mathematical background necessary for our proofs. This include some basic results on Sobolev functions, the quadratic form construction of $H_N(v,w)$, and some recent results concerning the ground-state of $H_N(v,w)$. 

The proofs of Theorems~\ref{thm:v-rep},~\ref{thm:v-rep non-local}, and~\ref{thm:v-rep non-interacting} are presented in Section~\ref{sec:v-rep}. In Section~\ref{sec:HK}, we turn to the proof of the Hohenberg-Kohn Theorems~\ref{thm:HK} and~\ref{thm:HK non-local}. In this section, we also present the proof of Proposition~\ref{prop:counterexample}. In Section~\ref{sec:xc}, we prove the last two main results, namely Theorems~\ref{thm:xc} and~\ref{thm:KS-DFT}.

In Section~\ref{sec:conclusion} we elaborate on possible extensions of the current results and related open questions. For the interested reader, we also present a formal derivation of the Kohn-Sham scheme in the current setting in Appendix~\ref{app:DFT}.

\section{Mathematical background on Sobolev spaces and quadratic forms}
\label{sec:background} 

In this section, we recall some well-known results about Sobolev spaces that will be useful throughout our proofs. We also recall the rigorous construction of the Schr\"odinger operator $H_N(v,w)$ via quadratic forms and present a few examples of distributional potentials. Lastly, we recall some recent results proven by the author concerning the ground-state of $H_N(v,w)$. These results will play a fundamental role throughout our proofs.

\subsection{Definition of Sobolev functions} We first recall the definition of Sobolev spaces.

\begin{definition}[Sobolev spaces]\label{def:Sobolev} For $1\leq p \leq \infty$, we denote by $\mW^{1,p}(I_N)$, the space of (complex-valued) functions $f\in \mL^p(I_N)$ with weak gradient $\nabla f \in \mL^p(I_N;\C^N)$ endowed with the norm
\begin{align}
    \norm{f}_{1,p}^p \coloneqq\norm{f}_{\mL^p}^p + \norm{\nabla f}_{\mL^p}^p.
\end{align}
Moreover, we denote by $\mW^{1,p}_0(I_N)$ the closure of the space $C^\infty_c(I_N)$ with respect to the $\mW^{1,p}$-norm. 
\end{definition}

\begin{definition}[Dual Sobolev spaces] \label{def:dualSobolev} For $1\leq p \leq \infty$, we denote by $\mW^{-1,p}(I_N)$, the dual space of $\mW^{1,q}(I_N)$, where $q$ is the H\"older conjugate of $p$ (i.e., $1/p+1/q = 1$) endowed with the operator norm. More precisely, 
\begin{align*}
    \mW^{-1,p}(I_N) \coloneqq \{ T: \mW^{1,q}(I_N) \rightarrow \C \quad \mbox{linear and continuous}\}\quad\mbox{with the norm}\quad \norm{T}_{-1,p} = \sup_{f\in \mW^{1,q} \setminus\{0\}} \frac{|T(f)|}{\norm{f}_{1,q}}.
\end{align*}
Similarly, we denote by $\mW^{-1,p}_0(I_N)$ the dual space of $W^{1,q}_0(I_N)$. 
\end{definition}

\begin{remark}[Notation]  For $p=2$, we use the standard notation $H^1(I_N)$ instead of $\mW^{1,2}(I_N)$. 
\end{remark}

Following the notation in \cite{Cor25b}, we define the space of periodic and anti-periodic functions as follows.
\begin{definition}[Sobolev Space with non-local BCs] Let $N\in \N$, then we denote by $\mH^1_{\pm 1}(I_N)$ the space of functions in $\mH^1(I_N)$ satisfying
\begin{align*}
    (\gamma\Psi)(x_1,...,x_{j-1},0,x_j,...,x_{N-1}) = \pm (\gamma \Psi)(x_1,....,x_{j-1},1,x_{j},...,x_{N-1})
\end{align*}
for almost every $(x_1,...x_{N-1}) \in I_{N-1}$ and any $j\in \{1,...,N\}$, where $\gamma : \mH^1(I_N) \rightarrow \mH^{1/2}(\partial I_N)$ denotes the standard Dirichlet trace operator\footnote{For the precise definition of $\mH^{1/2}(\partial \Omega)$ and the trace operator, we refer to \cite{EG15,Leo17}}.
\end{definition}
\begin{remark}[Non-local BCs plus anti-symmetry] \label{rem:symmetry+non-local} Note that, for antisymmetric functions, the periodicity (or anti-periodicity) condition only needs to be verified on one variable. More precisely, $\Psi \in \mH^1_{\pm 1}(I_N) \cap \mathcal{H}_N$ if and only if $\Psi \in \mH^1(I_N) \cap \mathcal{H}_N$ and
\begin{align*}
    (\gamma \Psi)(0,x') = \pm (\gamma \Psi)(1,x') \quad \mbox{for almost every $x'\in I_{N-1}$.}
\end{align*}
\end{remark}

\subsection{Regularity of reduced densities}

In this section, we recall some regularity results for the reduced densities of wave-functions with finite kinetic energy. 

To this end, and for later reference, let us recall the Gagliardo-Nirenberg-Sobolev (GNS) inequality. We refer, e.g., to \cite[Theorem 12.83]{Leo17} for a proof of the general $1\leq p,q \leq \infty$ case.

\begin{lemma}[GNS interpolation inequality]\label{lem:GNS} Let $\Omega \subset \R^d$ be a bounded open and connected domain with Lipschitz boundary. Then for any $f\in \mH^1(\Omega)$ and $2\leq p <\infty$ such that $\theta = d/2-d/p \in [0,1]$ we have 
\begin{align}
	\norm{f}_{\mL^p} \lesssim \norm{f}_{\mH^1}^{\theta} \norm{f}_{\mL^2}^{1-\theta} \quad\mbox{where}\quad \theta = \frac{d}{2} - \frac{d}{p}.\label{eq:GNS}
\end{align}
For $d=1$, the case $p =\infty$ is also allowed.
\end{lemma}

From the GNS inequality, one can prove, see \cite[Section 3]{Cor25b}, the following result. 
\begin{lemma}[Regularity of reduced densities] \label{lem:regularity reduced densities} Let $\Psi,\Phi \in \mH^1(\Omega \times \Omega')$ with $\Omega \subset \R^d$ and $\rho_{\Psi,\Phi}$ be the overlapping density
\begin{align}
\rho_{\Psi,\Phi}(x) \coloneqq \int_{\Omega'} \overline{\Psi(x,y)} \Phi(x,y) \mathrm{d} y. \label{eq:general density}
\end{align}
Then we have
\begin{align}
	\norm{\rho_{\Psi,\Phi}}_{1,p} \lesssim \norm{\Psi}_{\mH^1} \norm{\Phi}_{\mH^1}^{d-\frac{d}{p}} \norm{\Phi}_{\mL^2}^{1-d+\frac{d}{p}} + \norm{\Phi}_{\mH^1} \norm{\Psi}_{\mH^1}^{d-\frac{d}{p}} \norm{\Psi}_{\mL^2}^{1-d+\frac{d}{p}},  \,\, \mbox{for any}\,\,\begin{cases} 1\leq p \leq 2, \,\, \mbox{if $d=1$,} \\
  1\leq p < 2, \,\, \mbox{if $d=2$,} \\
  1 \leq p \leq \frac{d}{d-1}, \,\, \mbox{if $d\geq 3$,} \label{eq:reduced Sobolev}
 \end{cases}
\end{align}
where the implicit constant depends on $p$, $\Omega$, and $\Omega'$, but is independent of $\Psi$ and $\Phi$.
\end{lemma}



\subsection{Schr\"odinger operators with distributional potentials} \label{sec:self-adjoint} We now recall the construction of the self-adjoint operator $H_N(v,w)$.

For this, note that estimate~\eqref{eq:reduced Sobolev} and Young's inequality yields
\begin{align}
    |v(\rho_{\Psi}) + w(\rho^{(2)}_{\Psi})| \leq \epsilon \norm{\Psi}_{\mH^1}^2 + C_\epsilon \norm{\Psi}_{\mL^2}^2, \quad \mbox{for any $\Psi \in \mH^1(I_N)$,} \label{eq:KLMNest}
\end{align}
and for any $\epsilon>0$ provided that the constant $C_\epsilon>0$ is large enough. Hence, from the celebrated KLMN theorem, we can construct a unique self-adjoint operator associated to the form
\begin{align}
    \mathfrak{a}_{v,w}(\Psi,\Phi) = \int_{I_N} \overline{\nabla \Psi(x)} \scpr \nabla \Phi(x) \mathrm{d} x + v(\rho_{\Psi,\Phi}) + w(\rho^{(2)}_{\Psi,\Phi}), \label{eq:quadratic form}
\end{align}
for any closed subspace of $\mH^1(I_N)\cap \mathcal{H}_N$ that is dense in $\mathcal{H}_N$. In particular, the periodic, anti-periodic, Neumann, and Dirichlet operators are well-defined. 
\begin{lemma}[Schr\"odinger operators with distributional potentials] \label{lem:operator def} Let $I = (0,1)$, $N\in \N$, $v\in \mathcal{V}$, $w\in \mathcal{W}$, and
\begin{align*}
    \mathcal{Q}_N = \mH^1(I_N) \cap \mathcal{H}_N, \quad \mathcal{Q}_N^\pm = \mH^1_{\pm 1}(I_N) \cap \mathcal{H}_N, \quad \mbox{and}\quad \mathcal{Q}_N^0 = \mH^1_0(I_N) \cap \mathcal{H}_N.
\end{align*}
Then the sesquilinear form in~\eqref{eq:quadratic form} restricted to any of these form domains is closed, symmetric and semi-bounded. In particular, there exists unique self-adjoint operators $H_N(v,w)$, $H_N^\pm(v,w)$ and $H_N^0(v,w)$ associated to $\mathfrak{a}_{v,w}$ with the respective form domains. Moreover, all of these operators are semi-bounded and have purely discrete spectrum.
\end{lemma}

\subsection{Examples of distributional potentials} We now list a few distributional potentials for which the construction in Lemma~\ref{lem:operator def} applies.
\begin{enumerate}[label=(\arabic*)]
    \item The Dirac's delta potential $\delta_{x_0}$ with $x_0 \in I$ belongs to $\mathcal{V}$ by the continuous embedding $\mH^1(I) \subset C(\overline{I})$. Note that $x_0 \in \partial I$ is also allowed. 
    \item The $\delta$-interaction potential\footnote{We remark that, for spinless fermions, the $\delta$-interaction is void as the pair density vanishes on the diagonal (due to anti-symmetry of the wave-function).} $w = \delta(x-y)$, defined via
    \begin{align*}
        w_\delta(\rho^{(2)}_{\Psi,\Phi}) = \int_I \rho^{(2)}_{\Psi,\Phi}(x,x) \mathrm{d} x.
    \end{align*}
    \item Standard potentials $v\in \mL^1(I)$ and $w\in \mL^1((-1,1))$, whose actions are defined via
    \begin{align*}
        v(\rho) = \int_I v(x) \rho(x) \mathrm{d} x \quad \mbox{and}\quad w(\rho^{(2)}) = \int_{I_2} w(x-y) \rho^{(2)}(x,y) \mathrm{d} x \mathrm{d}y,
    \end{align*}
    are also allowed. 
    \item Lemma~\ref{lem:operator def} can also be extended to the case of $k^{th}$-body distributional potentials for $k \geq 3$. More precisely, for any $w_k\in \mW^{-1,q}(I^k)$ with $q>k$, one can use Lemma~\ref{lem:regularity reduced densities} to show that the form
    \begin{align*}
        \mathfrak{a}_{w_1,w_2,...,w_k}(\Psi,\Psi) \coloneqq \int_{I_N} |\nabla \Psi|^2\mathrm{d} x + \sum_{k=1}^N w_{k}(\rho_{\Psi}^{(k)}),
    \end{align*}
    where
    \begin{align*}
        \rho_{\Psi,\Phi}^{(k)}(x_1,..,x_k) = \frac{N!}{(N-k)! k!} \int_{I^{N-k}} \overline{\Psi(x_1,...,x_N)} \Phi(x_1,...,x_N)\mathrm{d}x_{k+1} ... \mathrm{d} x_N,
    \end{align*}
    is closed, symmetric and semibounded on any closed subspace of $\mH^1(I_N)$. Therefore, any Schr\"odinger operator of the form
    \begin{align*}
        -\Delta + \sum_{k=1}^N \sum_{j_1\neq ... \neq j_k}^N w_k(x_{j_1},x_{j_2},...,x_{j_k}), 
    \end{align*}
    with real-valued $w_k \in \mW^{-1,q_k}(I^k)$, where $q_k >k$ for $k\geq 2$ and $w_1 \in \mathcal{V}$, defines a semibounded self-adjoint operator with discrete spectrum. 
    \end{enumerate}

\subsection{Non-degeneracy and unique continuation property of the ground-state}

We end this section by recalling three results from \cite{Cor25b} that will play an important role in our proofs. 

The first result is that the ground-state of $H_N(v,w)$ is non-degenerate and satisfies the strong unique continuation property (cf. \cite[Theorems 2.1 and 2.3]{Cor25b}).

\begin{theorem}[Non-degeneracy theorem] \label{thm:non-degenerate}
    Let $v\in \mathcal{V}$, $w\in \mathcal{W}$ and $N\in \N$, then the operator $H_N(v,w)$ has a unique (up to a global phase) normalized ground-state $\Psi$ and $\Psi \neq 0$ almost everywhere in $I_N$. Moreover, the same result holds for $H_N^+(v,w)$, respectively $H_N^-(v,w)$, provided that the number of particles $N$ is odd, respectively even.
\end{theorem}

The second result is a weak unique continuation result along the boundary (cf. \cite[Theorem 6.1]{Cor25b}). 
\begin{theorem}(Weak unique continuation along the boundary) \label{thm:UCPboundary} Let $v\in \mathcal{V}$, $w\in \mathcal{W}$ and $N\in \N$, then the ground-state $\Psi$ of $H_N(v,w)$ can not vanish identically on a relatively open subset of the boundary $\partial I_N$. Moreover, if $N$ is odd, respectively even, then the same holds for the ground-state of $H_N^+(v,w)$, respectively $H_N^-(v,w)$. 
\end{theorem}

The last result shows that every eigenvalue of the single-particle operator $h(v) = -\Delta +v$ is at most two-fold degenerate and almost everywhere non-vanishing \cite[Theorem 2.6]{Cor25b}.  

\begin{theorem}[Spectrum of single-particle operator] \label{thm:single-particle} Let $v\in \mathcal{V}$ and $h^\pm(v) = -\Delta +v$ be the self-adjoint realizations of $\mathfrak{a}_{v,w}$ with domain $\mathcal{Q}_1^\pm = \mH^1_{\pm 1}(I)$. Then every eigenvalue of $h(v)$ is at most two-fold non-degenerate and every eigenfunction vanishes at most on a set of measure zero.
\end{theorem}

\begin{remark}[Finite vanishing set] In fact, one can use Courant's nodal domain theorem (see Lemma~\ref{lem:Courant}) to show that the eigenfunction associated with an eigenvalue $\lambda$ vanishes at most on $n(\lambda) = \sum_{\mu \leq \lambda} \mathrm{dim} \ker \left(\mu - h(v)\right)$ points.
\end{remark}

\section{Characterization of $v$-representable densities} \label{sec:v-rep}

Our goal in this section is to prove Theorems~\ref{thm:v-rep},~\ref{thm:v-rep non-local}, and~\ref{thm:v-rep non-interacting}. For the proof of Theorem~\ref{thm:v-rep}, we proceed in two steps. First, we show that any density in the set 
\begin{align}
    \mathcal{D}_N = \left\{ \rho \in \mH^1(I): \int_I \rho(x) \mathrm{d} x =N, \quad \rho(x) >0 \quad \mbox{for any $x\in [0,1]$}\right\} \label{eq:v-rep densities}
\end{align}
is $\mathcal{V}$-representable. Then we show that every $\mathcal{V}$-representable density belongs to $\mathcal{D}_N$. The proof of Theorem~\ref{thm:v-rep non-local} is also divided in these two steps. In fact, the proof of the first step is essentially identical as in the Neumann case, so we deal with all three BCs simultaneously in Section~\ref{sec:v-rep sufficient}. The second step requires some modifications, so we first address the Neumann case in Section~\ref{sec:v-rep local} and then the non-local BCs case in Section~\ref{sec:v-rep non-local}. The proof of Theorem~\ref{thm:v-rep non-interacting} is presented in Section~\ref{sec:v-rep non-interacting}.

\subsection{Sufficient conditions}  \label{sec:v-rep sufficient} In this section, we shall prove the following result.
\begin{lemma}[Sufficient conditions for $\mathcal{V}$-representability] \label{lem:sufficient v-rep} Let $\rho \in \mathcal{D}_N$, then for any $w\in \mathcal{W}$, there exists $v\in \mathcal{V}$ such that $\rho = \rho_{\Psi}$ where $\Psi$ is the ground-state wave-function of $H_N(v,w)$. Moreover, if $N$ is odd, respectively, even, then for any $\rho \in \mathcal{D}_N^+$, there exists $v\in \mathcal{V}$ such that $\rho = \rho_{\Psi}$ where $\Psi$ is the ground-state wave-function of $H_N^+(v,w)$, respectively, $H_N^-(v,w)$.
\end{lemma}

To prove this result, we adapt the approach of \cite{SPR+24}. Hence, let us introduce the space 
\begin{align}
    \mathcal{X}_N \coloneqq \left\{\rho \in \mH^1(I;\R): \int \rho(x) \mathrm{d} x = N\right\}. \label{eq:X space}
\end{align}
As $\mathcal{X}_N$ is an affine subspace of $\mH^1(I;\R)$, we can identify its dual with
\begin{align*}
    \mathcal{V}/\{1\} = \{ [v] : v\sim w \mbox{ if and only if $v-w=$ constant}\}.
\end{align*}
More precisely, we have $\mathcal{X}_N = N + \mathcal{X}_0 = \{ N+ \rho: \rho \in \mathcal{X}_0\}$ and $\mathcal{V}/\{1\} \cong (\mathcal{X}_0)^\ast$ via the dual pairing
\begin{align*}
    [v](\rho) = v(\rho) , \quad \mbox{for any $\rho \in \mathcal{X}_0$ and $v\in [v]$.}
\end{align*}

Next, let us define the space of density matrices with finite kinetic energy as
\begin{align}
    \mathcal{M}_N \coloneqq \left\{ \Pi = \sum_{j\geq 1} \mu_j P_{\Psi_j}: \inner{\Psi_j,\Psi_k}_{\mL^2} = \delta_{jk}, \quad 0 \leq \mu_j\leq 1 , \quad \sum_{j \geq 1} \mu_j = 1, \quad\mbox{and}\quad \sum_{j\geq 1} \mu_j \norm{\Psi_j}_{\mH^1}^2 <\infty \right\}, \label{eq:densitymatrices}
\end{align}
where $P_{\Psi_j}$ denotes the $\mL^2$-orthogonal projection on $\Psi_j$ and $\delta_{jk}$ is the Konecker's delta. The reduced densities of $\Pi$ are defined as
\begin{align*}
    \rho_{\Pi}(x) = \sum_{j} \mu_j \rho_{\Psi_j}(x) \quad \mbox{and}\quad \rho^{(2)}_\Pi(x,y) = \sum_{j} \mu_j \rho^{(2)}_{\Psi_j}(x,y).
\end{align*}
To simplify the notation we also define the kinetic energy of $\Pi$ as 
\begin{align}
    T(\Pi) = \sum_{j\geq 1} \mu_j \norm{\nabla \Psi_j}_{\mL^2}^2.  \label{eq:kinetic DM}
\end{align}

We now recall the following characterization of the set of $N$-representable densities, due to Lieb \cite{Lie83}. As the proof is rather short, we briefly sketch it below.

\begin{lemma}[$N$-representable densities] \label{lem:N-rep} Let $\Psi \in \mathcal{H}_N \cap \mH^1(I_N)$ be normalized, then $\rho_{\Psi}$ belongs to the set
\begin{align}
    \mathcal{R}_N \coloneqq \left\{ \rho : \sqrt{\rho} \in \mH^1(I), \quad \int_I \rho(x) \mathrm{d} x = N, \quad \mbox{and}\quad \rho(x) \geq 0 \quad\mbox{for any $x\in [0,1]$}\right\}. \label{eq:N-rep densities}
\end{align}
Conversely, if $\rho \in \mathcal{R}_N$, then there exists a Slater determinant $\Psi \in \mathcal{H}_N \cap \mH^1(I)$ such that $\rho_{\Psi} = \rho$ and 
\begin{align}
    \int_{I_N} |\nabla \Psi(x)|^2 \mathrm{d} x \lesssim 1 + \norm{\sqrt{\rho}}_{\mH^1}^2, \label{eq:kineticbound}
\end{align}
with an implicit constant independent of $\rho$. In the case of periodic and anti-periodic boundary conditions, i.e., if $\Psi \in \mH^1_{+1} \cap \mathcal{H}_N$ or $\Psi \in \mH^1_{-1}(I_N) \cap \mathcal{H}_N$, then the analogous result holds with $\mathcal{R}_N$ replaced by
\begin{align*}
    \mathcal{R}_N^{+} \coloneqq \mH^1_{+1}(I_N) \cap \mathcal{R}_N.
\end{align*}
\end{lemma}

\begin{proof} The first statement is a straightforward application of the Cauchy-Schwarz inequality. For the second part, one can define the functions
\begin{align}
    \phi_k(x) \coloneqq \sqrt{\frac{\rho(x)}{N}} \ee^{-\ii 2\pi  k F(x)}, \quad \mbox{where} \quad F(x) \coloneqq \frac{1}{N}\int_0^x \rho(y) \mathrm{d} y \quad \mbox{for $1\leq k \leq N$.} \label{eq:orbitaldef}
\end{align}
Then from the change of variables formula, we see that the functions $\{\phi_j\}_{j=1}^N$ are orthonormal in $\mL^2(I)$. In particular, the Slater determinant
\begin{align*}
    \Psi(x_1,...,x_N) = (\phi_1\wedge.. \phi_N)(x_1,...,x_N) = \frac{1}{\sqrt{N!}} \sum_{\sigma \in \mathcal{P}_N} \mathrm{sgn}(\sigma) \prod_{j=1}^N \phi_j(x_{\sigma(j)})
\end{align*}
is normalized and satisfies $\rho_{\Psi} = \rho$. Moreover, it holds that $\int |\nabla \Psi|^2 \mathrm{d} x = \sum_{j=1}^N \norm{\nabla \phi_j}_{\mL^2}^2$; therefore, the result follows from the inequality
\begin{align*}
    \int_I |\nabla \phi_j(x)|^2 \mathrm{d} x \lesssim \int_I
    \rho(x)^{3} + |\nabla \sqrt{\rho(x)}|^2 \mathrm{d} x \leq \norm{\rho}_{\mL^1} \norm{\sqrt{\rho}}_{\mL^\infty}^4 + \norm{\nabla \sqrt{\rho}}_{\mL^2} \lesssim_N 1 + \norm{\nabla \sqrt{\rho}}_{\mL^2}^2,
\end{align*}
where the last inequality follows from the GNS inequality~\eqref{eq:GNS} applied to $\sqrt{\rho}$ and the identity $\norm{\sqrt{\rho}}_{\mL^2}^2 = \norm{\rho}_{\mL^1} = N$.

In the case of periodic boundary conditions, the exact same argument gives the result. Indeed, if $\rho$ is periodic and integrates to $N$, then the orbitals $\phi_j$ defined in~\eqref{eq:orbitaldef} are automatically periodic as well. In the case of anti-periodic boundary conditions, one can replace the phase factor $\ii 2\pi k F(x)$ in~\eqref{eq:orbitaldef} by $\ii \pi k F(x)$ with $k$ odd. In this way, the orbitals are still orthonormal but anti-periodic.
\end{proof}

\begin{remark}[Only Neumann] For the remainder of this section, we shall work only in the Neumann case. The reason is that, by replacing the corresponding density and wave-function spaces by their periodic/anti-periodic counterparts, the exact same arguments lead to the corresponding results in these cases.
\end{remark}

Using Lieb's characterization of the set of $N$-representable densities, we can now introduce the relaxed constrained-search functional $F(\cdot;w) : \mH^1(I;\R) \rightarrow \R \cup \{+\infty\}$ as 
\begin{align*}
    F(\rho;w) = \begin{dcases} \min_{\Pi \in \mathcal{M}_N \rightarrow \rho} \{ T(\Pi) + w(\rho^{(2)}_{\Pi})\}
    , \quad &\mbox{for $\rho \in \mathcal{R}_N$,}\\
    +\infty, \quad &\mbox{otherwise.} \end{dcases}
\end{align*}
The following properties of $F$ were proved in \cite{Lie83} for regular (Coulomb) interaction potentials in $\R^N$. Here we provide a proof in the distributional case and for the bounded domain $I_N$.
\begin{lemma}[Constrained search over density matrices]\label{lem:Lieb2} For any $w\in \mathcal{W}$, the function $\rho \mapsto F(\rho;w)$ is convex. Moreover, for any $\rho \in \mathcal{R}_N$, the minimum in $F(\rho;w)$ is attained, i.e., there exists $\Pi_\rho \in \mathcal{M}_N$ such that $\Pi_\rho \mapsto \rho$ and $F(\rho;w) = T(\Pi_\rho) + w(\rho^{(2)}_{\Pi_\rho})$.
\end{lemma}
\begin{proof}
    To see that $F(\cdot;w)$ is convex, we first recall that $\mathcal{M}_N$ can be characterized as
    \begin{align}
        \mathcal{M}_N = \{ \Pi \in \mathcal{B}(\mL^2(I_N)): 0 \leq\Pi \leq 1, \quad\mathrm{Tr} \, \Pi = 1,\quad \mathrm{Tr} (H_N(0,0) \Pi) < \infty\},  \label{eq:densitymatrices2}
    \end{align}
    wher $\mathrm{Tr}$ denotes the trace of an operator. In particular, the set $\mathcal{M}_N$ is convex. Moreover, note that the linear map $\Pi \in \mathcal{M}_N \mapsto \rho_\Pi \in \mathcal{R}_N$ is surjective by Lemma~\ref{lem:N-rep}. Consequently, $\mathcal{R}_N$ is also a convex set. The convexity of $\rho \mapsto F(\rho;w)$ now follows from the fact that the map $\Pi \mapsto T(\Pi) + w(\rho^{(2)}_{\Pi}) = \mathrm{Tr} (H_N(0,w) \Pi)$ is linear.
    
    For the second statement, we use the direct method. This part of the proof is inspired by \cite{DFM08}. First, we let $\Pi_n = \sum \mu_{j,n} P_{\Psi_{j,n}}$ be a minimizing sequence of $T(\Pi_n) + w(\rho^{(2)}_{\Pi_n})$ satisfying $\rho_{\Pi_n} = \rho$. Then by the KLMN estimate~\eqref{eq:KLMNest} for $w$ only, there exists $C>0$ such that
    \begin{align}
        T(\Pi_n) &= \sum_{j \geq 1} \mu_{j,n} \norm{\nabla \Psi_{j,n}}_{\mL^2(I_N)}^2 \lesssim \sum_{j\geq 1} \mu_{j,n} \left(\norm{\nabla \Psi_{j,n}}_{\mL^2(I_N)}^2 + w(\rho^{(2)}_{\Psi_{j,n}}) + C \norm{\Psi_{j,n}}_{\mL^2(I_N)}^2\right) \nonumber \\
        &\leq T(\Pi_n) + w(\rho^{(2)}_{\Pi_n}) + C. \label{eq:unif bound}
    \end{align}
Thus $T(\Pi_n)$ is uniformly bounded in $n$. Since $0 \leq \mu_{j,n} \leq 1$ for any $n$, up to a subsequence we have, $\mu_{j,n}\rightarrow \mu_j$ as $n\ra \infty$ for some $0 \leq \mu_j \leq 1$. Therefore, for each $j$ such that $\mu_j>0$, the sequence $\{\Psi_{j,n}\}_{n \in \N}$ is bounded in $\mH^1(I_N)$. In particular, we can extract a weakly converging subsequence $\Psi_{j,n} \rightarrow \Psi_j$. Moreover, up to relabeling, we can assume that $\mu_j$ is ordered in non-increasing order. Hence, we can assume $\mu_j =0$ for any $j \geq M$, where $M$ could be finite or infinite. We then set
\begin{align*}
    \Pi_\rho \coloneqq \sum_{j=1}^M \mu_j P_{\Psi_j}.
\end{align*}
Then, for any $k\in \N$,
\begin{align*}
    \sum_{j=1}^k \mu_j = \sum_{j=1}^k \lim_n \mu_{j,n} \leq \lim_n \sum_{j=1}^\infty \mu_{j,n} = 1
\end{align*}
and, from~\eqref{eq:unif bound},
\begin{align*}
    \sum_{j=1}^k \mu_j \norm{\nabla \Psi_j}_{\mL^2} \leq \sum_{j=1}^k \liminf_n \mu_{j,n} \norm{\nabla \Psi_{j,n}}_{\mL^2} \leq \sup_{n} T(\Pi_n) < \infty.
\end{align*}
Passing to the limit $k\ra M$, we conclude that $0\leq \mathrm{Tr}\, \Pi_\rho \leq 1$ and
\begin{align}
    T(\Pi_\rho) \leq \liminf_n T(\Pi_n). \label{eq:kinetic}
\end{align}

We now claim that for any $\epsilon>0$ there exists $K>0$ such that
\begin{align}
    \sum_{j\geq K} \mu_{j,n} < \epsilon ,\quad \mbox{for any $n\in \N$.} \label{eq:claim2}
\end{align}
To prove this claim, note that from the variational principle we have
\begin{align}
    \sum_{j=1}^\ell \norm{ \nabla \Phi_j}_{\mL^2}^2 \geq \sum_{j=1}^\ell \lambda_j, \quad \mbox{for any $\mL^2$-orthonormal family $\{\Phi_j\}_{j=1}^\ell \subset \mL^2(I_N)$,} \label{eq:nice VP}
\end{align}
where $\lambda_j$ are the eigenvalues of the Neumann Laplacian on $I_N$ ordered increasingly. Moreover, from Weyl's law (or straightforward computations in the case of the hypercube $I_N$), we know that
\begin{align}
    \sum_{j=1}^\ell \lambda_j \gtrsim \ell^{1+\frac{2}{N}}. \label{eq:Weyl}
\end{align}
In particular, as the $\mu_{j,n}$ are ordered in non-increasing order, by~\eqref{eq:nice VP} and~\eqref{eq:Weyl} we obtain
\begin{align*}
    T(\Pi_n) \geq \sum_{j=1}^\ell \mu_{j,n} \norm{\nabla \Psi_{j,n}}_{\mL^2}^2 \geq \mu_{\ell,n} \sum_{j=1}^\ell \norm{\nabla\Psi_{j,n}}_{\mL^2}^2 \geq \mu_{\ell,n} \sum_{j=1}^\ell \lambda_j \gtrsim \mu_{\ell,n} \ell^{1+\frac{2}{N}}, 
\end{align*}
with an implicit constant independent of $n$. In particular, $\mu_{j,n} \leq C j^{-1-\frac{2}{N}}$ with constant $C>0$ independent of $n$. The claim now follows by choosing $K>0$ such that $\sum_{j\geq K}^\infty C j^{-1-\frac{2}{N}}<\epsilon$. 

From~\eqref{eq:claim2}, it is not hard to see that $\sum_{j=1}^M \mu_j =1$ and therefore $\Pi_\rho \in \mathcal{M}_N$. To conclude the proof, we now need to show that
\begin{align}
    \rho^{(2)}_{\Pi_n} \rightharpoonup \rho^{(2)}_{\Pi_\rho} \quad \mbox{ in $\mW^{1,p}(I_2)$.} \label{eq:W1p convergence}
\end{align}
Indeed, if this holds, then by~\eqref{eq:kinetic} we have $T(\Pi_\rho) + w(\rho^{(2)}_{\Pi_\rho}) \leq\liminf \{T(\Pi_n) + w(\rho^{(2)}_{\Pi_n})\} = F(\rho;w)$, which completes the proof. To prove~\eqref{eq:W1p convergence}, first note that, by inequality~\eqref{eq:reduced Sobolev} and H\"older's inequality
\begin{align*}
    \bignorm{\sum_{j\geq K} \mu_j \rho^{(2)}_{\Psi_j}}_{\mW^{1,p}} \leq \sum_{j\geq K} \bignorm{\rho^{(2)}_{\sqrt{\mu_j} \Psi_j}}_{\mW^{1,p}} \leq \sum_{j \geq K} \norm{\sqrt{\mu_j} \Psi_j}_{\mL^2}^{\frac{p}{2}} \norm{\sqrt{\mu_j} \Psi_j}_{\mH^1}^{2-\frac{p}{2}} \leq \left(1+T(\Pi)\right)^{1-\frac{p}{4}} \left(\sum_{j\geq K} \mu_j \right)^{\frac{p}{4}}.
\end{align*}
Thus by~\eqref{eq:claim2}, the weak convergence in~\eqref{eq:W1p convergence} follows if we can show that $\mu_{j,n} \rho^{(2)}_{ \Psi_{j,n}} \rightharpoonup \mu_j \rho^{(2)}_{\Psi_j}$ in $\mW^{1,p}$ for any $j\in \N$. Moreover, since $\lim \mu_{j,n} = \mu_j$, we just need to show that $\rho^{(2)}_{\Psi_{j,n}} \rightharpoonup \rho^{(2)}_{\Psi_j}$ in $\mW^{1,p}$ for $j\leq M$. For this, we rewrite 
\begin{align*}
    \rho^{(2)}_{\Psi_j} - \rho^{(2)}_{\Psi_{j,n}} =  \rho^{(2)}_{\Psi_j-\Psi_{j,n},\Psi_j} -\rho^{(2)}_{\Psi_j-\Psi_{j,n}} + \rho^{(2)}_{\Psi_j,\Psi_{j,n}-\Psi_j}. 
\end{align*}
Since $\Psi_{j,n} \rightarrow \Psi_j$ in $\mL^2$, we see from estimate~\eqref{eq:reduced Sobolev} that the middle term converges strongly (in $\mW^{1,p}(I_2)$) to zero. As the map $\Phi \mapsto \rho^{(2)}_{\Psi,\Phi}$ is linear, strong continuity from $\mH^1$ to $\mW^{1,p}$ implies weak continuity between the same spaces, and therefore, $\rho^{(2)}_{\Psi_j-\Psi_{j,n},\Psi_j} \rightharpoonup 0$ in $\mW^{1,p}$, which concludes the proof.
\end{proof}

Using the previous lemma, we obtain the following sufficient criteria for $\mathcal{V}$-representability.

\begin{lemma}[Criteria for $\mathcal{V}$-representability] \label{lem:v-rep criteria} A function $\rho \in \mathcal{R}_N$ is the (pure) ground-state density of an operator of the form $H_N(v,w)$ if and only if the subgradient of the function $\rho \mapsto F(\rho;w)$ is non-empty at $\rho$. Moreover, in this case, the minimum in $F$ is attained by the density matrix of the unique ground-state $\Psi_\rho \in \mathcal{H}_N \cap \mH^1(I_N)$ of $H_N(v,w)$ and
\begin{align*}
    F(\rho;w) = \min_{\Pi \rightarrow \rho} \{ T(\Pi) + w(\rho^{(2)}_{\Pi}) \} = \min_{\substack{ \Psi \mapsto \rho}} \{\norm{\nabla \Psi}_{\mL^2}^2 + w(\rho^{(2)}_{\Psi}) \}= F_{\rm LL}(\rho_{\Psi};w) = \norm{\nabla \Psi_\rho}_{\mL^2}^2 + w(\rho^{(2)}_{\Psi_\rho}) .
\end{align*}
\end{lemma}

\begin{proof} First, recall that by definition, a functional $[-v]\in \mathcal{V}\setminus \{1\}$ belongs to the space of subgradients of $F$, denoted here by $\partial F(\rho)$, if and only if $F(\rho)<\infty$ and
\begin{align*}
    0\leq F(\rho+\delta) - F(\rho) +[v](\delta) = F(\rho+\delta) -F(\rho) + v(\rho+\delta) - v(\rho),\quad \mbox{for any $\delta \in \mathcal{X}_0$ and $v\in [v]$.}
\end{align*}
Therefore, $[-v] \in \partial F(\rho_0;w)$ implies that
\begin{align*}
    F(\rho_0;w) +v(\rho_0) &= \min_{\rho \in \mathcal{X}_N} \{F(\rho;w) + v(\rho)\} 
 = \min_{\rho \in \mathcal{R}_N} \{\inf_{\substack{\Pi \mapsto \rho \\ \Pi \in \mathcal{M}_N}} \left\{ T(\Pi) +w(\rho^{(2)}_{\Pi}) \} + v(\rho) \right\} = \min_{\substack{\Psi \in \mathcal{Q}_N\\ \norm{\Psi}=1}} \mathfrak{a}_{v,w}(\Psi,\Psi).
\end{align*}
Since a minimizer $\Pi_0$ of $F(\rho_0;w)$ exists by Lemma~\ref{lem:Lieb2}, we have
\begin{align*}
    T(\Pi_0) + w(\rho^{(2)}_{\Pi_0}) + v(\rho_{0}) = F(\rho_0;w) = \min_{\substack{\Psi \in \mathcal{Q}_N\\ \norm{\Psi}=1}} \mathfrak{a}_{v,w}(\Psi,\Psi).
\end{align*}
Hence, $\Pi_0$ is an ensemble ground-state of $H_N(v,w)$. Since the ground-state of $H_N(v,w)$ is non-degenerate by Theorem~\ref{thm:non-degenerate}, we conclude that $\Pi_0$ is the density matrix of the unique ground-state. In particular, $\rho_0$ is the density of the unique ground-state of $H_N(v,w)$, which concludes the proof.
\end{proof}
Therefore, in order to complete the proof of Lemma~\ref{lem:sufficient v-rep}, it suffices to show that $\partial F(\rho;w) \neq \emptyset$ for any $\rho \in \mathcal{D}_N$. For this, we first note that the following inclusions holds:
\begin{align*}
    \mathcal{D}_N \subset \mathcal{R}_N \subset \mathcal{X}_N.
\end{align*}
Indeed, the first inclusion follows from the simple estimate
\begin{align}
    \norm{\nabla \sqrt{\rho}}_{\mL^2}^2 = \int_I \frac{|\nabla \rho(x)|^2}{\rho(x)} \mathrm{d} x \leq \norm{\nabla \rho}_{\mL^2}^2 \norm{1/\rho}_{\mL^\infty}, \label{eq:est1}
\end{align}
while the second inclusion follows from 
\begin{align}
    \int_I |\nabla \rho(x)|^2 \mathrm{d}x \leq \norm{\nabla \sqrt{\rho}}_{\mL^2}^2 \norm{\rho}_{\mL^\infty} \lesssim \norm{\sqrt{\rho}}_{\mL^2}^2 (\norm{\rho}_{\mL^1} + \norm{\nabla \sqrt{\rho}}_{\mL^2}^2), \label{eq:second inclusion}
\end{align}
where we used the GNS inequality~\eqref{eq:GNS} in the case $d=1$. Moreover, the GNS inequality also implies that the set $\mathcal{D}_N$ is open in $\mathcal{X}_N$. In fact, the following slightly stronger statement holds.
\begin{lemma}[Interior of $N$-representable densities] \label{lem:open} Let $\mathcal{D}_N, \mathcal{R}_N$, and $\mathcal{X}_N$ be defined via~\eqref{eq:v-rep densities},~\eqref{eq:N-rep densities}, and~\eqref{eq:X space}. Then 
\begin{align*}
    \mathcal{D}_N = \mathrm{int} \, \mathcal{R}_N \subset \mathcal{X}_N,
\end{align*}
where the interior is taken with respect to the $\mH^1$ topology on $\mathcal{X}_N$.
\end{lemma}
\begin{proof}
    That $\mathcal{D}_N$ is relatively open in $\mathcal{X}_N$ follows from the GNS inequality~\eqref{eq:GNS} (case $d=1$). For the other inclusion, let $\rho \in \mathcal{R}_N \setminus \mathcal{D}_N$, then $\rho(x) = 0$ for some $x \in [0,1]$. We can now take any $\delta \in \mathcal{X}_0$ such that $\delta(x)>0$ and note that $\rho_\epsilon\coloneqq \rho -\epsilon \delta \not \in \mathcal{R}_N$ for any $\epsilon>0$ because $\rho_\epsilon(x)<0$.
\end{proof}

The last ingredient we need for the proof of Lemma~\ref{lem:sufficient v-rep} is the following abstract result from convex analysis. The proof of this result can be found in several standard references, see, e.g. \cite[Proposition 5.2]{ET99}. 

\begin{lemma}[Existence of subgradient] \label{lem:subgradient} Let $F: X \rightarrow \R \cup\{+\infty\}$ be a convex functional in a locally convex topological vector space $X$. If $F$ is bounded on a neighboorhod of $\rho \in \mathrm{dom}\, F$, then $F$ is continuous at $\rho$ and $\partial F(\rho) \neq \emptyset$.
\end{lemma}

\begin{proof}[Proof of Lemma~\ref{lem:sufficient v-rep}]
According to Lemmas~\ref{lem:v-rep criteria} and~\ref{lem:subgradient}, it suffices to show that $F$ is uniformly bounded on a neighborhood of $\rho \in \mathcal{D}_N$. For this, simply note that, since $\mathcal{D}_N$ is relatively open in $\mathcal{X}_N$ (Lemma~\ref{lem:open}), for any $\rho \in \mathcal{D}_N$ we can find $\epsilon = \epsilon(\rho)>0$ such that 
\begin{align*}
    B_\epsilon(\rho) \coloneqq \{\rho' \in \mathcal{X}_N : \norm{\rho'-\rho}_{\mH^1} < \epsilon\} \subset \mathcal{D}_N \subset \mathcal{R}_N.
\end{align*}
Moreover, if we choose $\epsilon>0$ small enough, by~\eqref{eq:est1} we have
\begin{align*}
    \norm{\nabla \sqrt{\rho'}}_{\mL^2} \leq C, \quad \mbox{for any $\rho' \in B_\epsilon(\rho)$.}
\end{align*}
 To conclude, for any $\rho' \in \mathcal{R}_N$, there exists a wave-function $\Psi_{\rho'}$ satisfying the kinetic energy bound~\eqref{eq:kineticbound}, and therefore
 \begin{align*}
     F(\rho';w) \leq \inner{\Psi_{
     \rho'} H_N(v,0) \Psi_{\rho'}} \lesssim \norm{\Psi_{\rho'}}_{\mH^1}^2 \lesssim (1+ \norm{\nabla \sqrt{\rho'}}_{\mL^2}) \lesssim C,
 \end{align*}
 which completes the proof.
\end{proof}

\subsection{Necessary conditions - Neumann case} \label{sec:v-rep local} The following lemma shows that the ground-state density of $H_N(v,w)$ for any $v\in \mathcal{V}$ and $w\in \mathcal{W}$ belongs to the set $\mathcal{D}_N$. This result together with Lemma~\ref{lem:sufficient v-rep} completes the proof of Theorem~\ref{thm:v-rep}.

\begin{lemma}[Necessary conditions for $\mathcal{V}$-representability] \label{lem:necessary v-rep} Let $N \in \N$, $v\in \mathcal{V}$, $w\in \mathcal{W}$, and $\Psi$ be the (normalized) ground-state of $H_N(v,w)$. Then the density $\rho_{\Psi}$ belongs to the set $\mathcal{D}_N$ defined in~\eqref{eq:v-rep densities}.
\end{lemma}

For the proof of this result, we shall use the following version of Courant's nodal domain theorem \cite{Cou23,CH89}.

\begin{lemma}[Courant's nodal domain theorem] \label{lem:Courant} Let $\Psi$ be an eigenfunction of $H_N(v,w)$ with eigenvalue $\lambda$. Let $\{U_j\}_{j\leq M}$ be a collection of nonempty disjoint open subsets of $I_N$ such that the functions
\begin{align}
    \Psi_j(x) \coloneqq \begin{cases} \Psi(x), \quad &\mbox{for $x\in U_j$,}\\
    0, \quad &\mbox{otherwise,} \end{cases} \label{eq:eigenfunction}
\end{align} 
are not identically zero and belong to $\mH^1(I_N) \cap \mathcal{H}_N$. Then $M \leq n(\lambda) = \sum_{\mu\leq \lambda} \mathrm{dim} \ker(\mu - H_N(v,w))$.
\end{lemma}

\begin{remark*}[Nodal domain] We avoid any mention to the \emph{nodal domains} of $\Psi$ in Lemma~\ref{lem:Courant} because the eigenfunctions of $H_N(v,w)$ are in general not continuous, and we are not aware of any reasonable definition of nodal domain for (purely) $\mH^1$ functions.
\end{remark*}

The proof of Lemma~\ref{lem:Courant} follows the exact same steps as in the proof of Courant's nodal domain theorem for Schr\"odinger operators with regular potentials. For convenience of the reader, we briefly sketch the proof below.

\begin{proof}[Proof of Lemma~\ref{lem:Courant}] Let $\Psi$ be an eigenfunction of $H_N(v,w)$ with eigenvalue $\lambda$ and let $\{U_j\}_{j \leq M}$ satisfy the hypothesis from Lemma~\ref{lem:Courant}. Let $\Psi_j \in \mH^1(I_N)\cap \mathcal{H}_N$ be the functions defined in~\eqref{eq:eigenfunction}. Since the $U_j$'s are all disjoint, we have $\Psi_j \Psi_k =0$ for $j\neq k$. In particular, they are linearly independent and satisfy
\begin{align*}
   \mathfrak{a}_{v,w}(\Psi_j,\Psi_k) =  \int_{I_N} \overline{\nabla \Psi_j(x)} \scpr \nabla \Psi_k(x)  \mathrm{d}x  + v(\rho_{\Psi_j,\Psi_k}) + w(\rho^{(2)}_{\Psi_j,\Psi_k}) =0 \quad \mbox{for $j \neq k$.}
\end{align*}
On the other hand, since $\Psi$ is an eigenfunction with eigenvalue $\lambda$, a similar calculation shows that
\begin{align*}
    \lambda\norm{\Psi_j}_{\mL^2}^2 = \lambda \inner{\Psi, \Psi_j}_{\mL^2} = \mathfrak{a}_{v,w}(\Psi,\Psi_j) = \mathfrak{a}_{v,w}(\Psi_j,\Psi_j) \quad \mbox{for any $1\leq j \leq M$.}
\end{align*}
Consequently, any $F = \sum f_j \Psi_j\in \mathrm{span}\{\Psi_j : j \leq M\}$ satisfies
\begin{align*}
    \mathfrak{a}_{v,w}(F,F) = \sum_{j,k=1}^M \mathfrak{a}_{v,w}(f_j \Psi_j, f_k \Psi_k) = \sum_{j=1}^M |f_j|^2 \mathfrak{a}_{v,w}(\Psi_j,\Psi_j) = \lambda \sum_{j=1}^M |f_j|^2 \norm{\Psi_j}_{\mL^2}^2  = \lambda \norm{F}_{\mL^2}^2.
\end{align*}
As $\mathrm{dim} \, \mathrm{span} \{\Psi_j : 1\leq 
j \leq M\} = M$, it follows from the min-max principle that
\begin{align*}
    \lambda_M = \inf_{\substack{V \subset \mH^1(I_N)\\ \mathrm{dim} V = M}} \max_{\Psi \in V\setminus\{0\}} \frac{\mathfrak{a}_{v,w}(\Psi,\Psi)}{\norm{\Psi}_{\mL^2}^2} \leq \lambda.
\end{align*}
In particular $M \leq \sum_{\mu \leq \lambda} \mathrm{dim} \ker (H_N(v,w) - \mu)$.
\end{proof}

We can now complete the proof of Lemma~\ref{lem:necessary v-rep} and hence the proof of Theorem~\ref{thm:v-rep}.

\begin{proof}[Proof of Lemma~\ref{lem:necessary v-rep}] From Lemma~\ref{lem:Lieb2}, we already know that $\sqrt{\rho} \in \mH^1(I)$ and $\int \rho \mathrm{d} x = N$. Moreover, we have shown in~\eqref{eq:second inclusion} that $\rho \in \mH^1(I)$. Thus, it suffices to show that $\rho(x) >0$ for any $x \in [0,1]$. 

To this end, first note that, by Theorem~\ref{thm:UCPboundary}, the ground-state $\Psi$ of $H_N(v,w)$ can not vanish  identically along the boundary faces $\{0\} \times I_{N-1}$ and $\{1\} \times I_{N-1}$. Hence 
\begin{align*}
    \rho_{\Psi}(x_1) = \int_{I_{N-1}} |\Psi(x_1,x_2,...,x_N)|^2 \mathrm{d} x_2...\mathrm{d} x_N >0 \quad \mbox{for $x_1 \in \{0,1\}$.}
\end{align*}
It remains to show that the density is positive inside the interval $I=(0,1)$. For this, let us argue by contradiction. Suppose that there exists $y\in I$ such that $\rho_{\Psi}(y) = 0$. Then, by antisymmetry, the trace of $\Psi$ vanishes along the union of hyperplanes 
\begin{align*}
    E_{y} \coloneqq \cup_{k=1}^N [0,1]^{k-1}\times \{y\} \times [0,1]^{N-k} .
\end{align*} 
In particular, if we define the sets 
\begin{align*}
    U_1(y) \coloneqq \{ (x_1,...,x_N) \in I_N : x_j < y \} \quad \mbox{and}\quad U_2(y) \coloneqq \{(x_1,...,x_N) \in I_N: x_j >y\},
\end{align*}
then the trace of the restrictions $\Psi \rvert_{U_j(y)}$ vanish on $\partial U_j(y) \cap I_N$. As these are Lipschitz sets, the functions
\begin{align*}
    \Psi_1(x) \coloneqq \begin{cases} \Psi(x) \quad &\mbox{for $x\in U_1$,}\\
    0 \quad &\mbox{otherwise} \end{cases} \quad\mbox{and}\quad \Psi_2(x) \coloneqq \begin{cases} \Psi(x), \quad &\mbox{for $x\in U_2$,}\\
    0\quad &\mbox{otherwise,} \end{cases}
\end{align*}
belong to $\mH^1(I_N)$. Moreover, these functions are antisymmetric because $\Psi$ is antisymmetric and the domains $U_1$ and $U_2$ are invariant under permutation of coordinates. Consequently, by Lemma~\ref{lem:Courant} and the non-degeneracy of the ground-state $\Psi$ in Theorem~\ref{thm:non-degenerate}, we must have either $\Psi_1 = 0$ or $\Psi_2 = 0$. This now contradicts the unique continuation property of the ground-state in Theorem~\ref{thm:non-degenerate}, and therefore yields a contradiction.
\end{proof}

\subsection{Necessary conditions - non-local BCs} \label{sec:v-rep non-local} In this section we shall prove the following lemma, which together with Lemma~\ref{lem:sufficient v-rep} completes the proof of Theorem~\ref{thm:v-rep non-local}.
\begin{lemma}[Necessary conditions for $\mathcal{V}$-representability - non-local BCs] \label{lem:necessary v-rep non-local} Let $v\in \mathcal{V}$, $w\in \mathcal{W}$ and suppose that $N\in \N$ is odd. Then the ground-state density $\rho$ of $H_N^+(v,w)$ belongs to the set $\mathcal{D}_N^+$ defined in~\eqref{eq:v-rep Per}. On the other hand, if $N$ is even, then the ground-state density of $H_N^-(v,w)$ belongs to $\mathcal{D}^+_N$.
\end{lemma}

For this proof, we shall use the following result. 

\begin{lemma}[Rearranging the domain] \label{lem:rearranging} Let $x_\ast \in (0,1)$. For $y\in \R$, let $[y]$ denote the unique element in $[0,1)$ such that $y-[y] \in \Z$. Then the map $G_{\pm} : \mH^1_{\pm 1}(I_N)\cap \mathcal{H}_N \rightarrow \mH^1_{\pm 1}(I_N) \cap \mathcal{H}_N$ defined as
\begin{align}
    (G_{\pm} \Psi)(x_1,...,x_N) \coloneqq \left(\pm 1\right)^{m(x_1,...,x_N)} \Psi\left([x_1+x_\ast],[x_2+x_\ast], ..., [x_N+x_\ast]\right), \label{eq:G transform}
\end{align}
where
\begin{align}
    m(x) \coloneqq \sum_{j=1}^N \left((x_j+x_\ast)-[x_j+x_\ast]\right) \label{eq:exponent function}
\end{align}
is an isometric isomorphism. 
\end{lemma}

\begin{figure}[ht!]
    \centering
    \includegraphics[scale=0.52]{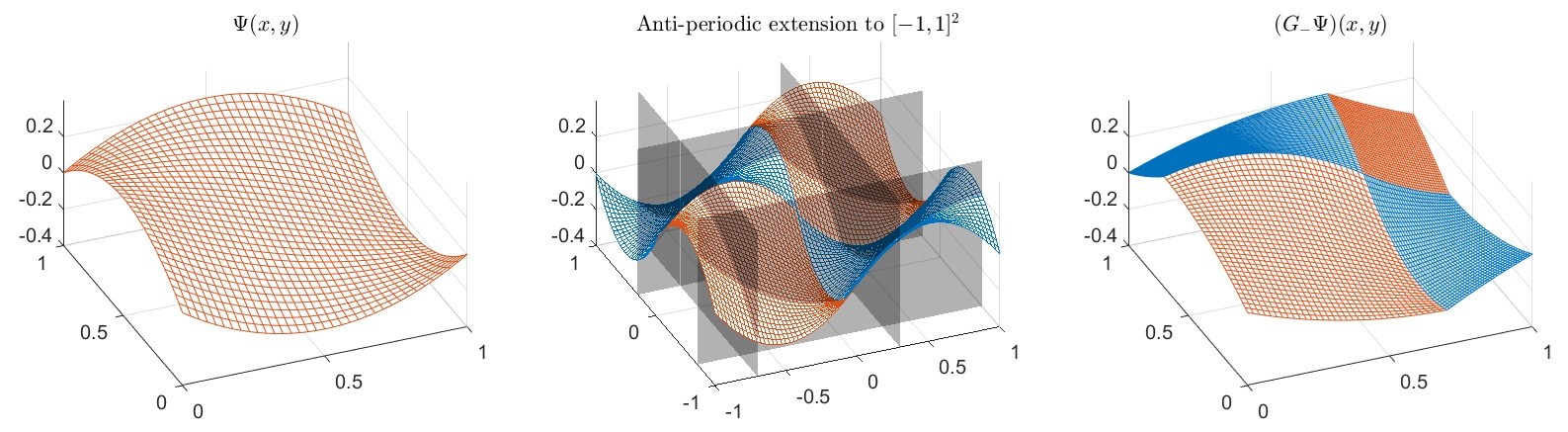}
    \caption{Visual illustration of the transformation $G_-$ with $N=2$ and $x_\ast = 1/3$. The middle plot shows the anti-periodic extension of $\Psi$ to $[-1, 1]^2$ with translations with different sign (in blue) and translations with same sign (in red). The planes $\{x= x_\ast\}, \{y=x_\ast\}, \{x=x_\ast -1\}$ and $\{y=x_\ast -1\}$, whose projections in the $(x,y)$-plane cover the boundary of the new box, are also depicted.}
    
 \label{fig:periodic_extension}
\end{figure}

\begin{proof}
    In the periodic case, the map $G_+$ is just a translation by $(x^\ast,....,x^\ast)$. Therefore, the result follows from the identification $\mH^1_{+1}(I_N) = \mH^1(\T^N)$, where $\T^N = \R^N/ \Z^N$ is the $N$-dimensional Torus. 
    
    For the anti-periodic case, a visual illustration of the transformation $G_-$ can be seen in Figure~\ref{fig:periodic_extension}. In this case, we first show that $G_-$ is an isomorphism from $\mH^1_{-1}(I_N)$ to $\mH^1_{-1}(I_N)$. For this, note that $G_- = G_1 \circ G_2 ... \circ G_N$, where
    \begin{align*}
        (G_j\Psi)(x_1,...,x_N) \coloneqq (-1) ^{x_j+x_\ast - [x_j+x_\ast]}\Psi(x_1,...,x_{j-1},[x_j+x_\ast], x_{j+1},...,x_N). 
    \end{align*}
    Thus it suffices to show that each $G_j$ is an isomorphism in  $\mH^1_{-1}(I_N)$. Let us show this for $G_1$. First, note that the restriction of $G_1 \Psi$ to $(0,1-x_\ast) \times I_{N-1}$ is just a translation of $\Psi$ restricted to $(x_\ast,1) \times I_{N-1}$. Similarly, the restriction of $G_1 \Psi$ to $(1-x_\ast, 1) \times I_{N-1}$ is the translation of $\Psi$ restricted to $(0,x_\ast) \times I_{N-1}$ times the constant $-1$. So clearly, $G_1 \Psi$ has a well-defined $\mL^2$-integrable gradient on each of the subdomains $(0,1-x_\ast) \times I_{N-1}$ and $(1-x_\ast, 1) \times I_{N-1}$ and satisfies
    \begin{align*}
        \norm{(G_- \Psi)}_{\mH^1((0,1-x_\ast) \times I_{N-1})}^2+\norm{(G_- \Psi)}_{\mH^1((1-x_\ast,1) \times I_{N-1})}^2 = \norm{\Psi}_{\mH^1(I_N)}^2.
    \end{align*}
    Moreover, as $\Psi$ is anti-periodic in $x_1$, we have
    \begin{align}
        \lim_{x_1 \uparrow 1-x_\ast} (G_1\Psi)(x_1,x') = \Psi(1,x') = -\Psi(0,x') = \lim_{x_1 \downarrow 1-x_\ast} (G_1 \Psi)(x_1,x'). \label{eq:funny}
    \end{align}
    Hence, the two pieces of $G_1 \Psi$ agree on $\{1-x_\ast\}\times I_{N-1}$ and therefore $G_1\Psi \in \mH^1(I_N)$. Similarly, by the definition of $G_1$, we have
    \begin{align*}
        (G_1 \Psi)(0,x') = \lim_{x_1 \downarrow 0} \Psi(x_\ast + x_1,x') = \Psi(x_\ast, x') =  \lim_{x_1 \uparrow 1^+} \Psi(x_\ast + x_1-1,x') = -(G_1\Psi)(1,x'),
    \end{align*}
    and therefore $G_1\Psi$ is anti-periodic in $x_1$.

    To complete the proof, it remains to show that $G_-$ maps $\mathcal{H}_N$ to itself. This is immediate from the fact that the map $(x_1,...,x_N) \mapsto ([x_1+x_\ast],...,[x_N+x_\ast])$ commutes with any coordinate permutation $\sigma$ and the exponent function $m$ in~\eqref{eq:exponent function} is symmetric, i.e., 
    \begin{align*}
        m(\sigma(x_1,...,x_N)) = m(x_1,...,x_N) \quad \mbox{for any $\sigma \in \mathcal{P}_N$.}
    \end{align*}
\end{proof}

\begin{remark}[Isomorphism on any periodic and anti-periodic Sobolev spaces] \label{rem:p-isometry} The proof of Lemma~\ref{lem:rearranging} shows that, for any $1\leq p\leq \infty$, $G_\pm$ is an isometric isomorphism from $\mW^{1,p}_{\pm}(I_N)$ to itself. In particular, the adjoint map, defined as
\begin{align*}
    (G_\pm^\ast F)(\Psi) \coloneqq F(G_{\pm}\Psi), \quad F\in \mW^{-1,q}(I_N), \Psi \in \mW^{1,p}(I_N),
\end{align*}
is an isomorphism from $\mW^{-1,p}(I_N)$ to itself. This fact will be used in the proof below.
\end{remark}
We can now proceed with the proof of Lemma~\ref{lem:necessary v-rep non-local}.
\begin{proof}[Proof of Lemma~\ref{lem:necessary v-rep non-local}] By contradiction, suppose that $\rho_{\Psi}(1-x_\ast) = 0$ for some $x_\ast \in (0,1)$. Then, we can define the operator $\widetilde{H} \coloneqq (G_{\pm})^{-1} H_N^\pm(v,w) G_{\pm}$. More precisely, $\widetilde{H}$ is the unique operator associated to the form
\begin{align*}
    \widetilde{a}_{v,w}(\Psi,\Phi) = \mathfrak{a}_{v,w}(G_{\pm} \Psi, G_{\pm} \Psi) = \int_{I_N} \overline{\nabla (G_\pm\Psi)(x)}\scpr \nabla (G_{\pm}\Phi)(x) \mathrm{d} x + v(\rho_{G_{\pm} \Psi,G_{\pm}\Phi}) + w(\rho^{(2)}_{G_{\pm}\Psi,G_{\pm} \Psi})
\end{align*}
The key observation now is that
\begin{align*}
    \overline{(G_{\pm}\Psi)(x)} (G_{\pm}\Phi)(x) = G_+\left(\overline{\Psi} \Phi\right)(x), \quad \mbox{for any $x\in I_N$.}
\end{align*}
From this observation, we see that
\begin{align*}
    \rho_{k,G_{\pm} \Psi,G_{\pm}\Phi}(x) = \int_{I_{N-k}} G_{+}(\overline{\Psi}\Phi)(x_1,...,x_N)\mathrm{d}x_{k+1}... \mathrm{d}x_N = (G_+ \rho_{k,\Psi,\Phi})(x_1,...,x_k),
\end{align*}
where the last $G_+$ denotes the version of $G_+$ acting on functions in $I_k$. Thus 
\begin{align*}
    v(\rho_{G_{\pm} \Psi,G_{\pm}\Phi}) + w(\rho^{(2)}_{G_{\pm}\Psi,G_{\pm} \Psi}) = \widetilde{v}(\rho_{\Psi,\Phi}) + \widetilde{w}(\rho^{(2)}_{\Psi,\Phi})
\end{align*}
where $\widetilde{v} = G_+^\ast v$ and $\widetilde{w} = G_+^\ast w$, and $G_{\pm}^\ast$ denotes the adjoint operator. From Remark~\ref{rem:p-isometry}, the adjoint operator is bounded from $\mW^{-1,p}(I_k)$ to $\mW^{-1,p}(I_k)$ for any $k\in \N$ and $1\leq p \leq\infty$; hence $\widetilde{v} \in \mathcal{V}$ and $\widetilde{w} \in \mathcal{W}$. As a consequence, 
\begin{align*}
    \widetilde{a}_{v,w}(\Psi,\Phi) = \int_{I_N} \overline{\nabla \Psi(x)}\scpr \nabla \Phi(x) \mathrm{d} x + \widetilde{v}(\rho_{\Psi,\Phi}) + \widetilde{w}(\rho^{(2)}_{\Psi,\Phi}) = a_{\widetilde{v},\widetilde{w}}(\Psi,\Phi).
\end{align*}
In other words, $\widetilde{H}= H^\pm_N(\widetilde{v},\widetilde{w})$. Since $G_{\pm}$ is an isometry, this operator is unitarily equivalent to $H^\pm(v,w)$. In particular, the ground-state $\widetilde{\Psi}$ of $H^\pm_N(\widetilde{v},\widetilde{w})$ is mapped to the ground-state $\Psi$ of $H^\pm_N(v,w)$ via $G_{\pm}$. Thus by~\eqref{eq:funny} 
\begin{align}
    (\pm 1)^{m(1-x_\ast, x')}\widetilde{\Psi}(0,[x_1'+x_\ast],....,[x_{N-1}'+x_\ast]) = (G_\pm \widetilde{\Psi})(1-x_\ast,x') = \Psi(1-x_\ast,x') \quad \mbox{for a.e. $x' \in I_{N-1}$.} \label{eq:move to boundary}
\end{align}  
Since we assumed $\rho_{\Psi}(1-x_\ast) = 0$, this implies that $\widetilde{\Psi}$ vanishes on the boundary face $\{0\} \times I_{N-1}$, which is not possible by Theorem~\ref{thm:UCPboundary}. Note that Theorem~\ref{thm:UCPboundary} is only proven under the parity assumption on the number of particles, which is why we need this assumption.
\end{proof}

\begin{remark}[Improved weak UCP along the boundary] The proof of Theorem~\ref{thm:UCPboundary2} follows from the argument above, i.e., from equation~\eqref{eq:move to boundary} by applying Theorem~\ref{thm:UCPboundary} to the ground-state $\widetilde{\Psi}$ of $H^\pm_N(\widetilde{v},\widetilde{w})$.
\end{remark}

\subsection{Characterization of non-interacting $\mathcal{V}$-representability for non-local BCS} \label{sec:v-rep non-interacting} We now turn to the proof of Theorem~\ref{thm:v-rep non-interacting}. For this proof, we shall use the following simple lemma.
\begin{lemma}[Convexity of the set of ground-state densities under two-fold degeneracy] \label{lem:convexity two-fold}Suppose that the ground-state of $H_N^\pm(v,w)$ is at most two-fold degenerate. Then the set
\begin{align*}
    \mathcal{D}^\pm_N(v,w) \coloneqq \{ \rho_{\Psi} : \Psi \mbox{ ground-state of $H_N^\pm(v,w)$}\}
\end{align*}
is convex. In particular, we have $\mathcal{D}^\pm_N(v,w) = \mathcal{D}^\pm_{N,\rm ens}(v,w)$, where
\begin{align*}
    \mathcal{D}_{N, \rm ens}^\pm(v,w) \coloneqq \left\{ \sum_{j\geq 1} t_j \rho_{\Psi_j} : 0\leq t_j \leq 1, \quad \sum_{j\geq 1} t_j = 1, \quad \{\Psi_j\} \mbox{ ground-states of $H_N^\pm(v,w)$}\right\}.
\end{align*}
\end{lemma}

\begin{proof} If the ground-state is non-degenerate, then the set $\mathcal{D}^\pm_N(v,w)$ has a single-point and the result trivially holds. If the ground-state is two-fold degenerate, we can find two orthogonal eigenfunctions. Moreover, since $H^\pm_N(v,w)$ commutes with complex conjugation, i.e., 
\begin{align*}
    \mathfrak{a}_{v,w}(\overline{\Psi}, \Phi) = \overline{\mathfrak{a}_{v,w}(\Phi, \overline{\Psi})} = \mathfrak{a}_{v,w}(\overline{\Phi}, \Psi),
\end{align*}
these eigenfunctions can be taken real-valued. Hence, let us denote by $\Psi_0$ and $\Psi_0'$ two orthogonal normalized real-valued ground-states, and by $\rho$ and $\rho'$ their single-particle densities. 

Now let $\rho_{\Psi_1}, \rho_{\Psi_2} \in \mathcal{D}_N^\pm(v,w)$. As the ground-state is two-fold degenerate, there exists $(\alpha_1,\beta_1),(\alpha_2,\beta_2)\in \C^2$ satisfying $|\alpha_j|^2 + |\beta_j|^2 = 1$ and $\Psi_j = \alpha_j \Psi_0 + \beta_j \Psi_0'$ for $j =1,2$. Hence
\begin{multline}
    (1-t)\rho_{\Psi_1} + t\rho_{\Psi_2} = \left((1-t) |\alpha_1|^2 + t |\alpha_2|^2\right) \rho + 2\left((1-t) \mathrm{Re}\,(\overline{\alpha_1}\beta_1) + t \mathrm{Re} \, (\overline{\alpha_2} \beta_2) \right) \rho_{\Psi_0,\Psi_0'} \\
    + \left((1-t)|\beta_1|^2 + t |\beta_2|^2 \right) \rho', \label{eq:weird}
\end{multline}
where $\rho_{\Psi_0,\Psi_0'}$ is the (real-valued) overlapping single-particle density of $\Psi_0$ and $\Psi_0'$. Define 
\begin{align}
    \alpha_3 \coloneqq \sqrt{(1-t)|\alpha_1|^2 + t|\alpha_2|^2} \quad \mbox{and}\quad \beta_3 \coloneqq \sqrt{(1-t)|\beta_1|^2 + t |\beta_2|^2}. \label{eq:alphadef}
\end{align}
Then, since by Cauchy-Schwarz 
\begin{align*}
    \left|(1-t) \mathrm{Re}\,(\overline{\alpha_1}\beta_1) + t \mathrm{Re} \, (\overline{\alpha_2} \beta_2) \right|^2 &\leq (1-t)^2 |\alpha_1 \beta_1|^2 + 2 t (1-t) |\alpha_1 \beta_1 \alpha_2 \beta_2| + t^2 |\alpha_2 \beta_2|^2 \\
    &\leq (1-t)^2 |\alpha_1 \beta_1|^2 + t(1-t) \left(|\alpha_1 \beta_2|^2 + |\alpha_2 \beta_1|^2\right) + t^2 | \alpha_2 \beta_2|^2 \\
    &= |\alpha_3 \beta_3|^2, 
\end{align*}
we can find $\theta \in [0,2\pi)$ such that
\begin{align}
    2\left((1-t) \mathrm{Re}\,(\overline{\alpha_1}\beta_1) + t \mathrm{Re} \, (\overline{\alpha_2} \beta_2) \right) = 2 \alpha_3 \beta_3 \cos(\theta) = 2 \mathrm{Re} (\alpha_3 \beta_3 \ee^{\ii \theta}). \label{eq:cosid}
\end{align}
Thus, by defining $\widetilde{\Psi} \coloneqq \alpha_3 \Psi_0 + \ee^{\ii \theta} \beta_3 \Psi_0'$ and using~\eqref{eq:weird},~\eqref{eq:alphadef}, and~\eqref{eq:cosid}, we find that
\begin{align*}
    \rho_{\widetilde{\Psi}} = \alpha_3^2 \rho_0 + 2\mathrm{Re} \, (\alpha_3 \beta_3 \ee^{\ii \theta}) \rho_{\Psi_0,\Psi_0'} + \beta_3^2 \rho_0' = (1-t) \rho_{\Psi_1} + t \rho_{\Psi_2}.
\end{align*}
As $\widetilde{\Psi}$ is also a normalized ground-state of $H_N^{\pm}(v,w)$, we conclude that $(1-t)\rho_{\Psi_1}+t \rho_{\Psi_2} \in \mathcal{D}_N^\pm(v,w)$. This shows that $\mathcal{D}_N^\pm(v,w)$ is convex.

Clearly we have $\mathcal{D}_N^\pm(v,w) \subset \mathcal{D}_{N,\rm ens}^\pm(v,w)$. Hence, to conclude the proof it suffices to prove the opposite inclusion. For this, let $\rho = \sum_{j} t_j \rho_{\Psi_j} \in \mathcal{D}^\pm_{N,\rm ens}(v,w)$ and observe that, if the sum has only finitely many terms, then $\rho \in \mathcal{D}_N^\pm(v,w)$ by the convexity of $\mathcal{D}^{\pm}_N(v,w)$. However, note that the ground-states $\Psi_j$ are not necessarily orthogonal to each other; in particular, there could be infinitely many of them. To overcome this issue, let us define $\Pi \coloneqq \sum_{j} t_j P_{\Psi_j}$, where $P_{\Psi_j}$ is the $\mL^2$-orthogonal projection on $\Psi_j$. Then, it is not hard to see that 
\begin{align*}
    0 \leq \Pi \leq 1, \quad \mathrm{Tr}\, \Pi = 1, \quad \mbox{and}\quad \mathrm{Tr}\, (H_N^\pm(0,0) \Pi) = \sum_{j} t_j \norm{\Psi_j}_{\mH^1}^2 \leq \sum_{j} t_j \left(\mathfrak{a}_{v,w}(\Psi_j,\Psi_j) + C\norm{\Psi_j}_{\mL^2}^2\right) < \infty.
\end{align*}
Consequently, by the spectral theorem there exists an $\mL^2(I_N)$-orthonormal family $\{\Phi_j\}_{j\geq 1} \subset \mathcal{Q}_N^{\pm}$ such that 
\begin{align}
    \Pi = \sum_{j} s_j P_{\Phi_j} \quad \mbox{where $0\leq s_j \leq 1$ and $\sum_{j} s_j = 1$.} \label{eq:sum0}
\end{align}
Since the $\Psi_j$'s are ground-states of $H_N^\pm(v,w)$ we have
\begin{align*}
    \lambda_1(v,w) 
 = \sum_{j} t_j \mathfrak{a}_{v,w}(\Psi_j,\Psi_j) = \mathrm{Tr} \; H_N^\pm(v,w) \Pi = \sum_{j} s_j \mathfrak{a}_{v,w}(\Phi_j,\Phi_j),
\end{align*}
where $\lambda_1(v,w)$ denotes the ground-state energy of $H_N^\pm(v,w)$. Therefore, each of the $\Phi_j$ is also a ground-state of $H_N^\pm(v,w)$. In particular $\rho_{\Phi_j} \in \mathcal{D}_N^\pm(v,w)$. Moreover, as there are at most two orthogonal ground-states (by assumption), the sum in~\eqref{eq:sum0} can run only up to $j=2$. Hence, 
\begin{align*}
    \sum_{j} t_j \rho_{\Psi_j} = \rho_{\Pi} = s_1 \rho_{\Phi_1} + (1-s_1) \rho_{\Phi_2} \in \mathcal{D}_N^+(v,w)
\end{align*}
by the convexity of $\mathcal{D}_N^\pm(v,w)$, which completes the proof.
\end{proof}

\begin{proof}[Proof of Theorem~\ref{thm:v-rep non-interacting}]
For simplicity, we present the proof only for the case of anti-periodic BCs. First note that, by the arguments in Section~\ref{sec:v-rep sufficient}, any density in the set $\mathcal{D}_N^+$ introduced in~\eqref{eq:v-rep Per} is ensemble $\mathcal{V}$-representable for any fixed interaction $w\in \mathcal{W}$. In particular, for $w =0$, we know that any density in $\mathcal{D}_N^+$ is non-interacting $\mathcal{V}$-representable, i.e., $\mathcal{D}_N^+ \subset \mathcal{D}_{N,\mathrm{ens}}^-(0)$, where
\begin{align*}
    \mathcal{D}_{N,\mathrm{ens}}^-(0) \coloneqq \left\{ \sum_{j\geq 1} t_j \rho_{\Psi_j} : 0 \leq t_j \leq 1 \quad \sum t_j = 1, \quad \Psi_j \mbox{ ground-state of $H_N^-(v,0)$ for some $v\in \mathcal{V}$}\right\}.
\end{align*}

Next, we observe that, since every eigenvalue of $h^-(v) = - \Delta +v$ is at most two-fold degenerate, the ground-state of the non-interacting operator $H_N^-(v,0)$ is at most two-fold degenerate. Therefore, by Lemma~\ref{lem:convexity two-fold}, every non-interacting ensemble $\mathcal{V}$-representable density is actually non-interacting pure-state $\mathcal{V}$-representable, or equivalently, $\mathcal{D}_N^-(0) = \mathcal{D}_{N,\mathrm{ens}}^-(0)$. In particular
\begin{align*}
    \mathcal{D}_{N}^+ \subset \mathcal{D}_{N}^-(0). 
\end{align*}
It remains to prove the opposite inclusion. For $N=2$,  this inclusion follows from case $w=0$ in Theorem~\ref{thm:v-rep non-local}. For $N\geq 3$, it suffices to show that $\rho_{\Psi}(x) >0$ for any ground-state $\Psi$ of $H_N^-(v,0)$ and any $x\in [0,1]$. For this, note that any ground-state of $H_N^-(v,0)$ can be written as
\begin{align*}
    \Psi = \phi_1 \wedge \phi_2 \wedge \Psi_{N-2},
\end{align*}
where $\phi_1,\phi_2$ are the first two eigenfunctions of $h^-(v)$ and $\Psi_{N-2}$ is a wave-function in the $N-2$-particle space $\mathcal{H}_{N-2}$ satisfying
\begin{align*}
    \int_{I} \phi_j(x) \Psi(x,x_2,...,x_{N-2}) \mathrm{d} x = 0\quad \mbox{for almost every $(x_2,...,x_{N-2}) \in I_{N-3}$.}
\end{align*}
This follows from the fact that the third eigenvalue of $h_-(v)$ is strictly larger than the second one by Theorem~\ref{thm:single-particle}, so the two lowest eigenfunctions have to be occupied in the ground-state. Hence, we have
\begin{align*}
    \rho_{\Psi} = |\phi_1|^2 + |\phi_2|^2 + \rho_{\Psi_{N-2}}.
\end{align*}
As $\Psi_2 \coloneqq \phi_1 \wedge \phi_2$ is the (unique by Theorem~\ref{thm:non-degenerate}) ground-state of $H_2^-(v,0)$, it follows from Theorem~\ref{thm:v-rep non-local} that $\rho_{\Psi_2}(x) = |\phi_1(x)|^2 + |\phi_2(x)|^2 >0$ and therefore $\rho_{\Psi}(x) >0$ for any $x\in [0,1]$. This proves that $\rho_{\Psi} \in \mathcal{D}_N^+$ and therefore the opposite inclusion holds.
\end{proof}

\section{The Hohenberg-Kohn theorem}\label{sec:HK}

In this section, we prove Theorems~\ref{thm:HK} and~\ref{thm:HK non-local}. We also present the proof of Proposition~\ref{prop:counterexample} and Theorem~\ref{thm:monotone}.

\subsection{Proof of the Hohenberg-Kohn theorem}

We start with the proof of Theorem~\ref{thm:HK}. 

\begin{proof}[Proof of Theorem~\ref{thm:HK}] Let $w\in \mathcal{W}$ and $N\in \N$ be fixed, and suppose that $H_N(v,w)$ and $H_N(v',w)$ for some $v,v'\in \mathcal{V}$ have the same ground-state density. Then from the standard Hohenberg-Kohn argument \cite{HK64} (see~\cite{PTC+23,Gar19}), we can show that both operators have a mutual ground-state wavefunction $\Psi$. Moreover, without loss of generality, we can assume that both ground-state energies are zero. In particular
\begin{align}
    0 = \mathfrak{a}_{v,w}(\Psi,\Phi) = \mathfrak{a}_{v',w}(\Psi,\Phi) = (v-v')(\rho_{\Psi,\Phi}) \quad \mbox{for any $\Phi \in \mH^1(I_N) \cap \mathcal{H}_N$.} \label{eq:HKeq}
\end{align}

The difficult part is to show that~\eqref{eq:HKeq} implies $v-v' =0$. To this end, let us define the following operator:
\begin{align}
    (Kf)(x) \coloneqq \int_I \frac{\rho^{(2)}_{\Psi}(x,y)}{\rho_{\Psi}(y)} f(y) \mathrm{d} y, \label{eq:T def}
\end{align}
where $\rho_{\Psi}$ and $\rho^{(2)}_{\Psi}$ are, respectively, the single-particle density and pair density of $\Psi$. The key observation here is that, due to the regularity of the pair density and the strict positivity of the density, the operator $K$ is regularity improving. More precisely, we have
\begin{lemma}[Regularity improving property]\label{lem:T operator} The operator $K$ defined above is a bounded linear operator from $\mL^\infty(I)$ to $\mH^1(I)$ and from $\mL^q(I)$ to $W^{1,p}(I)$ for any $1\leq p <2$ where $1=\frac{1}{q} + \frac{1}{p}$, i.e., 
\begin{align}
    &\norm{K f}_{\mW^{1,p}} \lesssim  \norm{f}_{\mL^q}, \quad \mbox{for any $f\in \mL^q(I)$ with $1=\frac{1}{q} + \frac{1}{p}$, and} \label{eq:pqregularity} \\
    &\norm{K f}_{\mH^1} \lesssim  \norm{f}_{\mL^\infty}, \quad \mbox{for any $f\in \mL^\infty(I)$,} \label{eq:infty regularity}
\end{align}
where the implicit constant is independent of $f$. 
\end{lemma}
\begin{proof}[Proof of Lemma~\ref{lem:T operator}]
Throughout this proof, let us abbreviate $\rho^{(2)}_{\Psi}$ and $\rho_\Psi$ by $\rho^{(2)}$ and $\rho$. Then first, by Lemma~\ref{lem:regularity reduced densities}, we have $\rho^{(2)} \in \mW^{1,p}(I_2)$ for any $1\leq p < 2$. Thus applying H\"older's inequality for the integral in $y$ we have
\begin{align*}
    \norm{Kf}_{\mL^p(I)} &= \left(\int_I \left|\int_I \rho^{(2)}(x,y) \frac{f(y)}{\rho(y)} \mathrm{d}y\right|^p \mathrm{d} x \right)^{\frac{1}{p}} \leq \norm{f/\rho}_{\mL^{q}} \norm{\rho^{(2)}}_{\mL^p} \leq \norm{f}_{\mL^{q}} \norm{1/\rho}_{\mL^\infty} \norm{\rho^{(2)}}_{\mL^p},
\end{align*}
where $\frac{1}{q} = 1- \frac{1}{p}$. Hence, the operator $K$ maps $\mL^{q}$ boundedly to $\mL^p$. Similarly, by replacing $\rho^{(2)}$ by $\partial_x \rho^{(2)}$ and using that $\partial_x \rho^{(2)} \in \mL^p(I_2)$ for any $1\leq p <2$ by Lemma~\ref{lem:regularity reduced densities}, we conclude that $K$ maps $\mL^q(I)$ to $\mW^{1,p}(I)$ for any $1\leq  p < 2$, which proves~\eqref{eq:pqregularity}.

To prove the estimate in~\eqref{eq:infty regularity}, we can use the fact that $\partial_x \rho^{(2)} \in \mL^2_x(I;\mL^1_y(I))$, i.e., 
\begin{align}
    \int_I \left|\int_I |\partial_x \rho^{(2)}(x,y)| \mathrm{d} y\right|^2 \mathrm{d} x < \infty. \label{eq:mixed ineq}
\end{align}
Indeed, assuming for the moment that~\eqref{eq:mixed ineq} holds, we have
\begin{align*}
    \norm{\partial_x (Kf)}_{\mL^2}^2 = \int_I \left| \int_I \partial_x \rho^{(2)}(x,y) \frac{f(y)}{\rho(y)} \mathrm{d} y\right|^2 \mathrm{d} x \leq \norm{f}_{\mL^\infty}^2 \norm{1/\rho}_{\mL^\infty}^2 \int_I \left(\int_I |\partial_x \rho^{(2)}(x,y)| \mathrm{d} y\right)^2 \mathrm{d} x 
\end{align*}
and therefore $K$ maps $\mL^\infty(I)$ to $\mH^1(I)$. Note that we can pass the derivative inside the integral in $\partial_x (Kf) = \int_I \partial_x \rho^{(2)}(x,y) (f/\rho)(y) \mathrm{d} y$ by approximating $\rho^{(2)}$ by smooth functions.

To prove~\eqref{eq:mixed ineq}, we first use Cauchy-Schwarz to obtain
\begin{align*}
    \left|\int_I |\partial_x \rho^{(2)}(x,y)| \mathrm{d} y\right|^2 &\leq N^2(N-1)^2\left(\int_{I_{N-1}} 2|\partial_x \Psi(x,y,x_3,...,x_N) \Psi(x,y,x_3,...,x_N)| \mathrm{d} x_3...\mathrm{d} x_N\mathrm{d} y \right)^2 \\
    &\leq 4 (N-1)^2 \rho_{\partial_x \Psi}(x) \rho_{\Psi}(x),
\end{align*}
where
\begin{align*}
    \rho_{\partial_x \Psi}(x) = N\int_{I_{N-1}} |\partial_x \Psi(x,x_2,...,x_N)|^2 \mathrm{d} x_2...\mathrm{d} x_N.
\end{align*}
Hence,~\eqref{eq:mixed ineq} follows from the simple estimates $\norm{\rho_{\partial_x \Psi}}_{\mL^1}\leq N\norm{\Psi}_{\mH^1}$ and $\norm{\rho_{\Psi}}_{\mL^\infty}\lesssim \norm{\Psi}_{\mH^1}$.
\end{proof}

The main idea now is to gain some regularity of $v-v'$ by testing~\eqref{eq:HKeq} with functions of the form 
\begin{align}
    \Phi_f(x_1,...,x_N) = N \sum_{j=1}^N \frac{f(x_j)}{\rho_{\Psi}(x_j)} \Psi(x_1,...,x_N) \quad \mbox{for $f \in \mH^1(I)$,} \label{eq:Phi_f}
\end{align}
and applying Lemma~\ref{lem:T operator}. Precisely, we note that, for any $f\in \mH^1(I)$, the function $\Phi_f$ defined above belongs to $\mathcal{H}_N \cap \mH^1(I_N)$ because $\norm{f}_{\mL^\infty} \leq \norm{f}_{\mH^1}, \norm{\rho_{\Psi}}_{\mL^\infty} \leq \norm{\Psi}_{\mH^1}$ and, since $\rho \in \mathcal{D}_N$, $\norm{1/\rho_{\Psi}}_{\mL^\infty} < \infty$. Moreover, the overlapping density of $\Phi_f$ and $\Psi$ is given by
\begin{align*}
    \rho_{\Psi,\Phi_f}(x) = N \int_{I_{N-1}} \overline{\Psi(x,x')} \Phi_f(x,x') \mathrm{d}x' = f(x) + \int_I \frac{\rho^{(2)}_{\Psi}(x,y)}{\rho_{\Psi}(y)} f(y) \mathrm{d} y = f(x) + (Kf)(x),
\end{align*}
where $K$ is defined in~\eqref{eq:T def}.
Hence, by plugging $\Phi_f$ in~\eqref{eq:HKeq} we obtain
\begin{align}
    (v-v')(f) + (v-v')(Kf) = 0, \quad \mbox{for any $f\in \mH^1(I)$.} \label{eq:HKid1}
\end{align}
By iterating this equation we have
\begin{align*}
    (v-v')(f) = (v-v')(K^2 f), \quad \mbox{for any $f\in \mH^1(I)$.}
\end{align*}
Thus from estimate~\eqref{eq:infty regularity}, the GNS inequality, and estimate~\eqref{eq:pqregularity}, we find that, for fixed $2 < q\leq  \infty$, 
\begin{align*}
    |(v-v')(f)| = |(v-v') (K^2 f)| \leq \norm{v-v'}_{\mH^{-1}} \norm{Kf}_{\mL^\infty} \leq \norm{v-v'}_{\mH^{-1}} \norm{Kf}_{\mW^{1,p}} \leq \norm{v-v'}_{\mH^{-1}} \norm{f}_{\mL^q}.
\end{align*}
As this holds for any $f\in \mH^1(I)$, which is dense in $\mL^q(I)$ for $2\leq q \leq \infty$, we can use the Riesz representation theorem in $\mL^q$ to conclude that $(v-v') \in \mL^p(I)$ for any $1\leq p <2$. 

Therefore~\eqref{eq:HKeq} implies that
\begin{align*}
    \sum_{j=1}^N (v-v')(x_j) \Psi(x_1,...,x_N) = 0 \quad \mbox{for almost every $(x_1,...,x_N) \in I_N$.}
\end{align*}
As $\Psi \neq 0$ a.e. by Theorem~\ref{thm:non-degenerate}, we find that
$\sum_{j=1}^N (v-v')(x_j) = 0$ for a.e. $(x_1,...,x_N) \in I_N$. Integrating all but one coordinate and using~\eqref{eq:HKeq}, we conclude that $v-v' = 0$, which completes the proof.
\end{proof}

We now sketch the modifications necessary to prove the Hohenberg-Kohn theorem for periodic and anti-periodic systems.
\begin{proof}[Proof of Theorem~\ref{thm:HK non-local}] We repeat the exact same steps as in the proof of Theorem~\ref{thm:HK} above. The only difference is that we can only test~\eqref{eq:HKeq} with periodic/anti-periodic functions $\Phi \in \mH^1_{\pm 1}(I_N) \cap \mathcal{H}_N$. Consequently, we can only take $\Phi_f$ as in~\eqref{eq:Phi_f} with $f \in \mH^1_{+1}(I)$, and therefore, ~\eqref{eq:HKid1} holds only for periodic $f$. Repeating the remaining arguments from the previous proof, we conclude that the restriction $(v-v') \rvert_{\mH^1_{+1}(I)}$ is regular (in $\mL^1(I)$), and therefore zero, which completes the proof.
\end{proof}

\subsection{Proof of Propositions~~\ref{prop:v-rep counterexample} and \ref{prop:counterexample} and Theorem~\ref{thm:monotone}}

We now turn to the proof of Propositions~\ref{prop:v-rep counterexample} and~\ref{prop:counterexample}.
\begin{proof}[Proof of Proposition~\ref{prop:v-rep counterexample}] From straightforward calculations, $\phi(x) = \sqrt{2}\cos(\pi x)$ is a ground-state of the anti-periodic Laplacian, hence $\rho(x) = 2\cos(\pi x)^2$ belongs to $\mathcal{D}^-_1$. As $\rho(1/2) =0$, it follows that $\mathcal{D}^-_1 \neq \mathcal{D}^+_1$ because the latter only contains strictly positive densities (by Theorem~\ref{thm:HK non-local}).
\end{proof}

\begin{proof}[Proof of Proposition~\ref{prop:counterexample}]
As $\sqrt{2}\cos(\pi x)$ is a ground-state of the anti-periodic Laplacian, we have
\begin{align*}
    \norm{\nabla \psi}_{\mL^2(I)}^2 \geq \pi^2 \norm{\psi}_{\mL^2(I)}^2 \quad \mbox{for any $\psi \in \mH^1_{-1}(I)$.}
\end{align*}
Therefore, clearly,
\begin{align*}
    \norm{\nabla \psi}_{\mL^2(I)}^2 + \alpha |\psi(1/2)|^2 \geq \pi^2 \norm{\psi}_{\mL^2(I)}^2, \quad \mbox{for any $\psi \in \mH^1_{-1}(I)$ and $\alpha>0$.}
\end{align*}
On the other hand, equality is achieved for $\psi(x) = \sqrt{2} \cos(\pi x)$. In particular, by the variational principle, $\psi(x) = \sqrt{2} \cos(\pi x)$ is a ground-state of $h^-(v) = -\Delta + \alpha \delta_{1/2}$ for any $\alpha>0$, which completes the proof.
\end{proof}

\begin{remark} There is nothing special about the point $x=1/2$. By considering functions of the form $\psi(x) = \sin(\pi x) - \beta \cos(\pi x)$, $\beta \in \R$, we can see that, for any $\alpha>$ and $x_0 \in [0,1]$, the anti-periodic realizations of $-\Delta + \alpha \delta_{x_0}$ and $-\Delta$ have a mutual ground-state density.
\end{remark}

We end this section with the proof of Theorem~\ref{thm:monotone}.

\begin{proof}[Proof of Theorem~\ref{thm:monotone}]
 Let $\Psi$ be the ground-state of $H_N(v,w)$. Then by the assumption $v-v' \geq 0$ and the variational principle we have
\begin{align*}
    \lambda_1(v,w) = \mathfrak{a}_{v,w}(\Psi,\Psi) \geq \mathfrak{a}_{v',w}(\Psi,\Psi) \geq \lambda_1(v',w).
\end{align*}
In particular, if the equality $\lambda_1(v,w) = \lambda_1(v',w)$ holds, then $\Psi$ is also a ground-state of $H_N(v'w,)$. By the HK theorem~\ref{thm:HK}, this implies that $v-v'$ is constant and therefore zero since $\lambda_1(v,w) = \lambda_1(v',w)$. Consequently, equality holds if and only if $v=v'$, which completes the proof. 
\end{proof}

\section{Differentiability of the exchange-correlation functional} \label{sec:xc} We now prove Theorem~\ref{thm:xc}. The first and main step in this proof is to show that the Levy-Lieb constrained-search functional is differentiable in $\mathcal{D}_N$. 
\begin{lemma}[Differentiability of the Levy-Lieb functional] \label{lem:LL} For any $w\in \mathcal{W}$ and $\rho \in \mathcal{D}_N$, the minimum in the Levy-Lieb constrained search functional
\begin{align}
    F_{\rm LL}(\rho;w) = \min_{\substack{\Psi \in \mathcal{Q}_N \\ \Psi \mapsto \rho}} \{\inner{\Psi, H_N(0,w) \Psi}\} \label{eq:LLfunctional0}
\end{align}
is attained by a unique (up to a global phase) wave-function $\Psi_\rho$. Moreover, $F_{\rm LL}$ is Gateaux-differentiable at any $\rho \in \mathcal{D}_N$ and the potential $v(\rho;w) = -\mathrm{d}_\rho F_{\rm LL}$  is the unique (up to an additive constant) potential such that $\rho$ is the ground-state density of $H_N\left(v(\rho;w),w\right)$.
\end{lemma}

\begin{proof}
    By Theorem~\ref{thm:v-rep} and Lemma~\ref{lem:v-rep criteria}, for any $\rho \in \mathcal{D}_N$ and $w\in \mathcal{W}$, there exists a $v_\rho \in \mathcal{V}$ such that $[-v_\rho] \in \partial F(\rho;w)$ and $\rho$ is the density of the ground-state $\Psi_\rho$ of $H_N(v_\rho,w)$. Moreover, Lemma~\ref{lem:v-rep criteria} also tell us that the minimum in $F(\rho;w)$ is uniquely achieved by the density matrix of the pure state $\Psi_\rho$. In particular, the minimum in $F_{\rm LL}(\rho;w)$ is achieved by the unique (up to a global phase) wave-function $\Psi_\rho$, which proves~\eqref{eq:LLfunctional0}.
    
    To prove the Gateaux differentiability of $F_{\rm LL}$, we first note that, from Theorem~\ref{thm:HK} and Lemma~\ref{lem:v-rep criteria}, there exists only one equivalence class $[-v] \in \mathrm{d}_\rho F$ for each $\rho \in \mathcal{D}_N$. Hence, the differentiability of $F$ follows from the following abstract result: any convex function $G:X \rightarrow \R\cup \{+\infty\}$ that is finite and continuous at a point $\rho \in X$ of a topological vector space $X$ is Gateaux differentiable at $\rho$ if and only if there exists a unique $v\in \partial G(\rho)$ (see, e.g., \cite[Proposition 5.3]{ET99}). The differentiability of $F_{\rm LL}$ then follows from the fact that $F_{\rm LL}(\rho;w) = F(\rho;w)$ for $\rho \in \mathcal{D}_N$ and this set is open in $\mathcal{X}_N$.
\end{proof}

In the next step, we show that the Kohn-Sham kinetic energy functional agrees with $F_{\rm LL}(\cdot;0)$. 
\begin{lemma}[Kinetic energy functional] \label{lem:KS kinetic} For any $\rho \in \mathcal{D}_N$, the unique minimizer in $F_{\rm LL}(\rho;0)$ is a Slater determinant. In particular,
\begin{align*}
    T_{\rm KS}(\rho) = F_{\rm LL}(\rho;0), \quad \mbox{for any $\rho \in \mathcal{D}_N$,}
\end{align*}
and $T_{\rm KS}$ is Gateaux-differentiable at any $\rho \in \mathcal{D}_N$.
\end{lemma}

\begin{proof}
   By Lemma~\ref{lem:LL}, the minimizer in $F_{\rm LL}(\rho;0)$ is given by a unique $\Psi_\rho$. Moreover, from Lemma~\ref{lem:LL} we also know that $\Psi_\rho$ is the ground-state of $H_N(v_\rho,0)$ for any $v_\rho$ in the equivalence class of $\mathrm{d}_\rho F_{\rm LL}(\cdot;0)$. As this ground-state is unique by Theorem~\ref{thm:non-degenerate}, it must be given by the Slater determinant of the $N$ lowest eigenfunctions of $-\Delta + v_\rho$, which shows that $T_{\rm KS}(\rho) = F_{\rm LL}(\rho;0)$ for any $\rho \in \mathcal{D}_N$. That $T_{\rm KS}$ is Gateaux-differentiable now follows from the fact that $\rho \mapsto F_{\rm LL}(\rho;0)$ is differentiable. 
\end{proof}

The next lemma shows that the Hartree functional is also differentiable.
\begin{lemma}[Differentiability of the Hartree functional] \label{lem:hartree} Let $w\in \mathcal{W}$, then the Hartree functional $E_H: \mH^1(I) \rightarrow \R$ given by
\begin{align*}
    \rho \mapsto E_H(\rho) = w(\rho\otimes \rho)
\end{align*}
is (Frechet) smooth and its derivative at $\rho \in \mH^1(I)$ is given by
\begin{align*}
    \delta \in \mH^1(I) \rightarrow \mathrm{d}_\rho E_H(\delta) = w(\rho \otimes \delta) + w(\delta \otimes \rho).
\end{align*}
\end{lemma}

\begin{proof}
    Since $E_H$ is the composition of the map $\rho \mapsto (\rho,\rho)$ with the bilinear map $(\rho,\delta) \rightarrow b(\rho,\delta) = w(\rho \otimes \delta)$, it suffices to show that $b$ is continuous from $\mH^1(I) \times \mH^1(I)$ to $\mW^{1,p}(I_2)$ for $1\leq p <2$. For this, note that
    \begin{align*}
        \norm{\rho \otimes \delta}_{\mH^1}^2 &= \norm{\rho \otimes \delta}_{\mL^2}^2 + \norm{\nabla (\rho\otimes \delta)}_{\mL^2}^2 = \norm{\rho \otimes \delta}_{\mL^2}^2 + \norm{(\partial_x \rho) \otimes \delta}_{\mL^2}^2 + \norm{\rho \otimes (\partial_x \delta)}_{\mL^2}^2\\
        &=\norm{\rho}_{\mL^2}^2 \norm{\delta}_{\mL^2}^2 + \norm{\partial_x \rho}_{\mL^2}^2\norm{\delta}_{\mL^2}^2 + \norm{\rho}_{\mL^2}^2 \norm{\partial_x \delta}_{\mL^2}^2 \leq \norm{\rho}_{\mH^1}^2 \norm{\delta}_{\mH^1}^2.
    \end{align*}
    Thus, from the continuous embedding $\mH^1(I_2) \hookrightarrow \mW^{1,p}(I_2)$ for $p<2$, we conclude that $b$ is continuous and therefore $E_H$ is smooth. The formula for the derivative follows from a straightforward computation.
\end{proof}

We can now complete the proof of Theorem~\ref{thm:xc}.

\begin{proof}[Proof of Theorem~\ref{thm:xc}] Since 
\begin{align}
    E_{\rm xc}(\rho;w) = F_{\rm LL}(\rho;w) - T_{\rm KS}(\rho) - E_H(\rho), \label{eq:xc0}
\end{align}
the map $\rho \mapsto E_{\rm xc}(\rho;w)$ is Gateaux-differentiable at any $\rho \in \mathcal{D}_N$ by Lemmas~\ref{lem:LL},~\ref{lem:KS kinetic}, and~\ref{lem:hartree}.
\end{proof}

We now present the proof of Theorem~\ref{thm:KS-DFT}. 

\begin{proof}[Proof of Theorem~\ref{thm:KS-DFT}] Since the ground-state of $H_N(v,w)$ exists and is unique, there exists a unique minimizer $\rho_0 = \rho(v,w)$ of
\begin{align*}
    E(v) &= \min_{\rho \in \mathcal{R}_N} \bigr\{\inf_{\substack{\Psi \in \mH^1(I_N) \cap \mathcal{H}_N\\
    \Psi \mapsto \rho}} \{\mathfrak{a}_{v,w}(\Psi,\Psi)\} \bigr\} = \min_{\rho \in \mathcal{R}_N} \{F_{\rm LL}(\rho;w) + v(\rho)\} \\
    &= \min_{\rho\in \mathcal{R}_N} \left\{T_{\rm KS}(\rho) + E_H(\rho;w) + E_{\rm xc}(\rho) + v(\rho)\right\}
\end{align*}
Moreover, $\rho_0 \in \mathcal{D}_N$ by Theorem~\ref{thm:v-rep}. In particular, by Lemma~\ref{lem:KS kinetic}, there exists a unique Slater determinant $\Psi_{\rm KS} \in \mathcal{S}_N$ such that $T_{\rm KS}(\rho_0) = \norm{\nabla \Psi_{\rm KS}}_{\mL^2}^2$ and $\rho_{\Psi_{\rm KS}} = \rho_0$. In particular, $\Psi_{\rm KS}$ is the unique (up to a global phase) minimizer of the problem 
\begin{align*}
    \min_{\Psi \in \mathcal{S}_N} \{ \norm{\nabla \Psi}_{\mL^2}^2 + E_H(\rho_\Psi;w) + E_{\rm xc}(\rho_\Psi) + v(\rho_\Psi)\} . 
\end{align*}

It remains to show that $\Psi_{\rm KS}$ is the Slater determinant of the $N$ lowest eigenfunctions of
\begin{align*}
   - \Delta + v_{\rm xc}(\rho_0) + v_H(\rho_0) + v\quad \mbox{where}\quad v_{\rm xc}(\rho_0) = \mathrm{d}_{\rho_0} E_{\rm xc} \quad \mbox{and}\quad v_H(\rho_0) = \mathrm{d}_{\rho_0} E_H.
\end{align*}
For this, note that by Lemma~\ref{lem:KS kinetic}, $\Psi_{\rm KS}$ is the unique minimizer of
\begin{align*}
    F_{\rm LL}(\rho_0;0) = \min_{\substack{\Psi \in \mH^1(I_N)\cap \mathcal{H}_N \\ \Psi \mapsto \rho_0}} \norm{\nabla \Psi}_{\mL^2}^2.
\end{align*}
Thus, by Lemmas~\ref{lem:LL} and~\ref{lem:KS kinetic}, $\Psi_{\rm KS}$ is the ground-state of 
\begin{align*}
    H_N(v_{\rm KS}(\rho_0),0), \quad \mbox{where} \quad v_{\rm KS}(\rho_0) \coloneqq -\mathrm{d}_{\rho_0} F_{\rm LL}(\cdot,0) = -\mathrm{d}_{\rho_0} T_{\rm KS}. 
\end{align*}
As the ground-state of $H_N(v_{\rm KS}(\rho_0),0)$ is non-degenerate by Theorem~\ref{thm:non-degenerate}, $\Psi_{\rm KS}$ is the Slater determinant of the $N$ lowest eigenfunctions of the single-particle operator 
\begin{align*}
    h_{\rm KS}(\rho_0) \coloneqq -\Delta + v_{\rm KS}(\rho_0).
\end{align*}
The result now follows from the identity
\begin{align*}
    v_{\rm KS}(\rho_0) = -\mathrm{d}_{\rho_0} T_{\rm KS} =  \mathrm{d}_{\rho_0}\left(E_{\rm xc} + E_H - F_{\rm LL}(\cdot,w)\right) = v_{\rm xc}(\rho_0) + v_H(\rho_0) + v,
\end{align*}
which is immediate from~\eqref{eq:xc0}.
\end{proof}

\section{Concluding remarks}
\label{sec:conclusion}

In this paper, we characterized the set of ground-state densities of many-body Schr\"odinger operators for spinless fermions living in a one-dimensional interval with Neumann boundary conditions. This gives a complete solution to the pure-state $v$-representability problem in this setting. Moreover, it shows that the set of $v$-representable densities is independent of the interaction potential. We then obtained a Hohenberg-Kohn theorem for distributional potentials in the class $\mathcal{V}$. In particular, these two results show that, for any fixed interaction $w$ in a large class of distributions, there exists a one-to-one correspondence between the set $\mathcal{D}_N$ and the set of external potentials $\mathcal{V}$ modulo additive constants. Furthermore, we proved that the exchange-correlation functional is differentiable and the exchange-correlation well defined. Combining these results, we established that the Aufbau principle holds and the Kohn-Sham scheme is rigorously exact. In other words, the ground-state density of any interacting systems of fermions in one dimension can be exactly reproduced via the Kohn-Sham scheme.  

In addition, we established analogous results in the case of periodic and anti-periodic boundary conditions. We also presented a counter example to the Hohenberg-Kohn theorem for distributional potentials in the case of anti-periodic BCs, which highlights the importance of BCs. To conclude, let us now comment on possible extensions of the current results and related open questions.
\begin{enumerate}[label=(\arabic*)]

\item ($\mathcal{V}$-representability in the Dirichlet case) While Theorem~\ref{thm:non-degenerate} applies to the case of Dirichlet BCs, it is not clear how to extend the convex analysis argument to provide sufficient conditions for $\mathcal{V}$-representability in this case. The main problem is that the set of Dirichlet $N$-representable densities has empty interior (with respect to the $\mH^1$-topology) in its affine hull. More precisely, the following holds.
\begin{proposition}[Empty interior in the Dirichlet case] \label{prop:no interior} Let $N\in \N$ and $\mathcal{R}_N^0$ be the set of $N$-representable densities with Dirichlet boundary conditions
\begin{align*}
    \mathcal{R}_N^0 = \left\{\rho : \sqrt{\rho} \in \mH^1_0(I), \quad \int_I \rho(x) \mathrm{d}x = N,\quad\mbox{and}\quad \rho \geq 0\right\}.
\end{align*} 
Then the relative interior of $\mathcal{R}^0_N$ in $\mathrm{aff}(\mathcal{R}_N^0)$ with respect to the $\mH^1_0$-topology is empty.
\end{proposition}
Nevertheless, we remark that the results established in \cite{Cor25b} give a hint on a further necessary condition for $\mathcal{V}$-representability in the Dirichlet case. More precisely, those results show that the Neumann trace of the ground-state wave-function is well-defined and nowhere vanishing along the boundary of $I_N$. Roughly speaking, this implies that the ground-state density $\rho$ should be proportional to $(x-x_0)^2$ as $x$ approaches an end point $x_0$ of the interval.

It is also not clear how to adapt the arguments to the case of unbounded intervals (e.g. $I = \R$) as, similar to the bounded Dirichlet case, the set of $N$-representable densities can be shown to have empty interior. Furthermore, in this case one has to deal with the additional difficulty that a ground-state is not guaranteed to exist. 
\item (Hohenberg-Kohn theorem in the Dirichlet case) The proof of the Hohenberg-Kohn theorem is not immediate to extend to the Dirichlet case. The difficulty in this case is that $1/\rho$ is no longer uniformly bounded. Nevertheless, it should be noted that $1/\rho$ is still locally bounded inside the interval. Therefore, we believe that the proof here can be adapted to this case and the analogous result from Theorem~\ref{thm:HK} holds.

\item (Spin electrons) Another natural open question is whether the current results can be extended to \emph{spin} electrons. Note that, in this case, the anti-symmetry of the wave-function is with respect to exchanging both spatial and spin coordinates simultaneously. Hence, one of the main ingredients in our proofs, the non-degeneracy Theorem~\ref{thm:non-degenerate} does not immediately applies. Nevertheless, we remark that this non-degeneracy theorem can be extended to the case of wave-functions with partial anti-symmetry, i.e., anti-symmetry with respect to exchanging only a subset of spatial coordinates. In particular, we speculate that all of the results presented here can be extended to the spin case. 

\item (Higher dimensions) It is not clear (at least to the author) how to extend the current results to electrons living in two or three dimensional spaces. First, in this case, no non-degeneracy theorem holds and a complete solution to the pure-state $\mathcal{V}$-representability problem seems out of reach. Second, it is not clear how to prove ensemble $v$-representability via the same convex analysis argument used here. More precisely, it is not clear what are the "correct" class of distributional potentials that one should consider to be able to represent a reasonable set of densities. Nevertheless, we remark that the class of potentials $\mathcal{V}$ considered here have an interesting mathematical characterization, namely, they can be identified with the class of all local and real-valued infinitesimal form perturbations of the one-dimensional Laplacian. While a more precise statement will appear only in a future contribution, we emphasize that an analogous characterization of all local form perturbations of the Laplacian in higher dimensions could lead to significant insights into the class of potentials to be considered for the $v$-representability problem.
\end{enumerate}

\addtocontents{toc}{\protect\setcounter{tocdepth}{1}}

\addtocontents{toc}{\protect\setcounter{tocdepth}{-1}}
\section*{Acknowledgements}
The author thanks Chokri Manai, Markus Penz, Sarina Sutter, Robert van Leuween, and Michael Ruggenthaler for several fruitful and inspiring discussions on the topic of this paper. The author is also grateful to Andre Laestadius and Vebjørn Bakkestuen for organizing the workshop on Foundations and Extensions of DFT which led to the ideas for the current paper. Finally, the author is also grateful to the anonymous referees for carefully examining the paper, spotting several typos, and making many helpful suggestions to improve the presentation.

T.C.~Corso acknowledges funding by the \emph{Deutsche Forschungsgemeinschaft} (DFG, German Research Foundation) - Project number 442047500 through the Collaborative Research Center "Sparsity and Singular Structures" (SFB 1481). 

\addtocontents{toc}{\protect\setcounter{tocdepth}{2}}
\appendix

\section{Proof of Proposition~\ref{prop:no interior}}

We now prove Proposition~\ref{prop:no interior}. 
\begin{proof}[Proof of Proposition~\ref{prop:no interior}] We first claim that
\begin{align}
    V\coloneqq \left\{\rho + f: \rho \in \mathcal{R}_N^0 , \quad f\in \mH^1_0(I) ,\quad \int_I f \mathrm{d} x = 0,\quad  \mathrm{supp}(f) \subset \subset I\right\} \subset \mathrm{Aff}(\mathcal{R}_N^0). \label{eq:claim3}
\end{align}
To see this, let $\rho_0 \in \mathcal{R}_N^0$ be such that $\rho_0(x)>0$ for any $x\in I$. Then for any $f\in \mH^1_0(I)$ with compact support in $I$ there exists $C>0$ such that $|f(x)| \leq C \rho_0(x)$. This follows from the fact that $c_0 \coloneqq \inf_{x\in \mathrm{supp}(f)} \rho_0(x) >0$ and $f\in \mL^\infty(I)$. Hence, $\rho_0 + f/(2C) \geq \rho_0/2 \geq 0$. Therefore
\begin{align*}
    \left|\partial_x \left(\sqrt{\rho_0+f/(2C)}\right)\right| = \left|\frac{\partial_x \rho_0 + \partial_x f/(2C)}{\sqrt{\rho_0+f/(2C)}}\right|\leq \frac{|\partial_x \rho_0|}{\sqrt{\rho_0/2}} + \frac{1}{2C} \frac{|\partial_x f|}{\sqrt{c_0/2}}
 \in \mL^2(I).\end{align*}
Thus $\sqrt{\rho_0 + f/(2C)} \in \mH^1_0(I)$. In particular, if $\int_I f = 0$, then $\rho_0+f/(2C) \in \mathcal{R}_N^0$, which shows that 
\begin{align}
    \left\{\rho_0 + f: f \in \mH^1_0(I), \quad \int_I f = 0 \quad \mathrm{supp}(f) \subset \subset I\right\} \subset \mathrm{Aff}(\mathcal{R}_N^0). \label{eq:some middle eq}
\end{align}
As $\mathrm{Aff}(\mathcal{R}_N^0)$ is affine, we can replace $\rho_0$ by any $\rho \in \mathcal{R}_N^0$ in~\eqref{eq:some middle eq}, which proves~\eqref{eq:claim3}.

We can now show that the relative interior of $\mathcal{R}_N^0$ in $\mathrm{Aff}(\mathcal{R}_N^0)$ is empty as follows. First, suppose that $\rho \in \mathcal{R}_N^0$ satisfies $\rho(x_0) = 0$ for some $x_0 \in I$. Then for any $\epsilon>0$ we can find $f_\epsilon \in C^\infty_c(I)$ such that $\norm{f_\epsilon}_{\mH^1} \leq \epsilon$ and $f_\epsilon(x_0) >0$. Consequently, $\rho-f_\epsilon \not \in\mathcal{R}_N^0$ but $\rho - f_\epsilon \in V \subset \mathrm{Aff}(\mathcal{R}_N^0)$. As $\epsilon>0$ is arbitrary, $\rho$ is not on the relative interior of $\mathcal{R}_N^0$ in $\mathrm{Aff}(\mathcal{R}_N^0)$. Now suppose that $\rho \in \mathcal{R}_N^0$ and $\rho(x)>0$ for any $x\in I$. Then by dominated convergence, for any $\epsilon>0$ we can find $0< \delta =\delta(\epsilon)<1/4$ such that
\begin{align*}
    0< \norm{\rho}_{\mH^1(0,\delta_\epsilon)} \leq \epsilon. 
\end{align*}
Moreover, as $\lim_{x\downarrow 0} \rho(x) = 0$ we can find $0<\delta' < \delta$ such that $\rho(\delta') < \rho(\delta)/2$. Now construct a function $f_\epsilon$ such that $f_\epsilon(x) = (\rho(x) - \rho(\delta'))_+$ for $\delta'\leq x\leq \delta$, $f_\epsilon(x) = 0$ for $x\leq \delta'$ or $x\geq 1-\delta'$, $\int f_\epsilon = 0$ and $\norm{f_\epsilon} \leq C \epsilon$ with a constant independent of $\epsilon>0$. Note that this is possible because $\norm{f_\epsilon}_{\mH^1(0,\delta)} \leq \norm{\rho_\epsilon}_{\mH^1(0,\delta)} \leq \epsilon$. Then, on the one hand, $\rho -2f_\epsilon \in  \mathrm{Aff}(\mathcal{R}_N^0)$ by~\eqref{eq:claim3}. On the other hand,
\begin{align*}
    \rho(\delta)-2f_\epsilon(\delta)=-\rho(\delta)+2\rho(\delta') < 0,
\end{align*}
and therefore $\rho + 2f_\epsilon \not \in \mathcal{R}_N^0$. As $\epsilon>0$ can be taken arbitrarily small, this concludes the proof. 
\end{proof}

\section{The Kohn-Sham Scheme}
\label{app:DFT}
In this section we briefly present a formal derivation of the Kohn-Sham (KS) scheme. 

    In most applications of DFT, one is mainly interested in computing the single-particle density of the ground-state of a system described by a Hamiltonian of the form $H_N(v,w)$ for a large number of electrons $N\in \N$. Hence, one seeks to solve the following ground-state problem: find the minimum and the minimizer\footnote{Here we have a minima in~\eqref{eq:gs problem} instead of infimum because a ground-state always exists (see Section~\ref{sec:self-adjoint}).} of
\begin{align*}
    E_{\rm GS}(v;w) \coloneqq \min_{\substack{\Psi \in \mathcal{Q}_N \\ \norm{\Psi} = 1}} \inner{\Psi, H_N(v,w) \Psi},
\end{align*}
where $\mathcal{Q}_N = \mathcal{H}_N \cap \mH^1(I_N)$ is the quadratic form domain of $H_N(v,w)$. However, due to the large spatial dimension of functions in $\mathcal{Q}_N$ when $N$ is large, solving the ground-state problem with standard variational methods is not feasible. 

To bypass this difficulty, the core idea of DFT is that the ground-state problem should be reformulated as a minimization problem in terms of the (single-particle) density only. As shown by Levy \cite{Lev79} and Lieb \cite{Lie83}, this can indeed be achieved by setting
\begin{align}
    E_{\rm GS}(v;w) = \min_{\rho \in \mathcal{R}_N} \{F_{\rm LL}(\rho;w) + v(\rho) \}, \label{eq:gs problem}
\end{align}
where $\mathcal{R}_N$ is the set of $N$-representable densities~\eqref{eq:N-rep densities} and $F_{\rm LL}$ is the celebrated Levy-Lieb constrained search functional
\begin{align*}
    F_{\rm LL}(\rho;w) \coloneqq \inf_{\Psi \in \mathcal{Q}_N \mapsto \rho} \inner{\Psi, H_N(0,w) \Psi}. 
\end{align*}
Here and henceforth the notation $\Psi \mapsto \rho$ means that the single-particle density of $\Psi$ is given by $\rho$. Unfortunately, no real gain is obtained via this reformulation as each evaluation of $F_{\rm LL}$ requires again a minimization over the high-dimensional space of wave-functions. 

To tackle the ground-state problem, Kohn and Sham \cite{KS65} proposed the following scheme. First, one defines the exchange-correlation functional as
\begin{align*}
    E_{\rm xc}(\rho;w) \coloneqq F_{\rm LL}(\rho;w) - T_{\rm KS}(\rho) - E_H(\rho;w), 
\end{align*}
where $E_H$ denotes the Hartree energy
\begin{align*}
    E_H(\rho;w) \coloneqq w(\rho \otimes \rho) 
\end{align*}
and $T_{\rm KS}$ is the Kohn-Sham kinetic energy functional
\begin{align*}
    T_{\rm KS}(\rho) = \min_{\substack{\Psi \in \mathcal{S}_N \\ \Psi \mapsto \rho}}  \int_{I_N} |\nabla \Psi(x_1,...,x_N)|^2 \mathrm{d} x_1 ... \mathrm{d} x_N.
\end{align*}
The minimization\footnote{The existence of a minimizer in~\eqref{eq:KS kinetic} can be shown via the direct method (cf. \cite[Theorem 4.7]{Lie83}), see also Lemma~\ref{lem:KS kinetic}.} in $T_{\rm KS}$ is over the set of Slater determinants with finite kinetic energy, i.e., the set
\begin{align*}
    \mathcal{S}_N \coloneqq \{\Psi = \phi_1 \wedge... \wedge \phi_N : \{\phi_j\}_{j=1}^N \subset \mH^1(I) \quad \mbox{and}\quad \inner{\phi_i, \phi_j}_{\mL^2(I)} = \delta_{ij} \quad \mbox{for $1\leq i,j\leq N$} \}.
\end{align*}

Then, notice that by construction,
\begin{align*}
    E_{\rm GS}(v;w) = \inf_{\rho \in \mathcal{R}_N} \left\{T_{\rm KS}(\rho) + E_H(\rho) + E_{\rm xc}(\rho) +v(\rho)\right\} = \inf_{\Psi \in \mathcal{S}_N } \{\norm{\nabla \Psi}_{\mL^2}^2 + E_H(\rho_{\Psi}) + E_{\rm xc}(\rho_{\Psi}) + v(\rho_{\Psi})\}.
\end{align*}
The key observation now is that the map $\vec{\phi} = (\phi_1,...,\phi_N) \mapsto S(\vec{\phi}) = \Psi = \phi_1 \wedge ... \wedge \phi_N$ is a smooth surjection from the (Grassmanian) manifold of orbitals
\begin{align*}
    \mathcal{G}_M \coloneqq \left\{ \vec{\phi} =(\phi_1,...,\phi_N) \in \left(\mH^1(I)\right)^N : \quad \inner{\phi_i,\phi_j}_{\mL^2} = \delta_{ij} , \quad \mbox{for $1\leq i,j\leq N$}\right\}
\end{align*}
to the space of Slater determinants $\mathcal{S}_N$; hence, the ground-state problem can be re-stated as
\begin{align*}
    E_{\rm GS}(v;w) = \inf_{\vec{\phi} \in \mathcal{G}_M} \left\{\mathcal{E}(\vec{\phi}) \coloneqq \sum_{j=1}^N \norm{\nabla \phi_j}_{\mL^2}^2 + (E_H+E_{\rm xc} + v)(\rho_{\vec{\phi}})\right\}, \quad \mbox{with}\quad \rho_{\vec{\phi}} \coloneqq \sum_{j=1}^N |\phi_j|^2.
\end{align*}

Therefore, under the assumption that $\rho \mapsto E_{\rm xc}(\rho;w)$ is differentiable with Gateaux derivative $\mathrm{d}_\rho E_{\rm xc} \in \mathcal{V}$, a vector $\vec{\phi} \in \mathcal{G}_M$ is a critical point of the above minimization problem if and only if it satisfies the Euler-Lagrange equation
\begin{align*}
    \sum_{j=1}^N \inner{\nabla \phi_j, \nabla \psi_j} + \sum_{k=1}^N (\mathrm{d}_{\rho_{\vec{\phi}}} E_H + \mathrm{d}_{\rho_{\vec{\phi}}} E_{\rm xc} + v)(\overline{\phi_j} \psi_j) = \sum_{i,j} \mu_{ij} \inner{\phi_j,\psi_k}, \quad \mbox{for any $\vec{\Psi} = (\psi_1,...,\psi_N) \in \mH^1(I)^N$,} 
\end{align*}
and some Lagrange multipliers $\mu_{ij}$. In fact, the Lagrange multipliers $\{\mu_{ij}\}_{i,j}$ can be shown to be a self-adjoint matrix. Hence, using the fact that the energy is invariant under unitary transformation of the orbitals, i.e., $\mathcal{E}(U \vec{\phi}) = \mathcal{E}(\vec{\phi})$ for any unitary $U\in \C^{N \times N}$, the previous equation is equivalent to
\begin{align}
   \inner{\nabla \phi_j, \nabla \psi} + (\mathrm{d}_{\rho_{\vec{\phi}}} E_{H} + \mathrm{d}_{\rho_{\vec{\phi}}} E_{\rm xc}+v)(\overline{\phi_j} \psi) = \lambda_j \inner{\phi_j,\psi} \quad \mbox{for any $\psi \in \mH^1(I)$ and $1\leq j, \leq N$.} \label{eq:KS}
\end{align}

In other words, $\vec{\phi}$ is a critical point of $\mathcal{E}$ if and only if, up to a unitary matrix $U\in \C^{N\times N}$, the orbitals $\phi_j$ are eigenfunctions of the Kohn-Sham single-particle Hamiltonian
\begin{align*}
    h_{\rm KS}(\rho_{\vec{\phi}}) \coloneqq -\Delta + v_H(\rho_{\vec{\phi}}) + v_{\rm xc}(\rho_{\vec{\phi}}) + v,
\end{align*}
where $v_H(\rho)$ is the Hartree (distributional) potential
\begin{align*}
   \delta \mapsto  v_H(\rho)(\delta) = w(\rho \otimes \delta) + w(\delta \otimes \rho), 
\end{align*}
and $v_{\rm xc}(\rho)$ is the so-called exchange-correlation potential
\begin{align*}
    \delta \mapsto v_{\rm xc}(\rho)(\delta) = \mathrm{d}_\rho E_{\rm xc}(\delta).
\end{align*}

\begin{remark*}[Aufbau principle]
For a minimizer of $\mathcal{E}$, the orbitals $\vec{\phi}\in \mathcal{G}_M$ are expected to be the lowest eigenfunctions of $h_{\rm KS}(\rho_{\vec{\phi}})$. This is the so-called \emph{Aufbau principle} and, though not entirely justified in general, can be shown to hold in the current setting (cf. Theorem~\ref{thm:KS-DFT}).
\end{remark*}

The equations in~\eqref{eq:KS} are called the \emph{Kohn-Sham equations} and are the fundamental equations in (Kohn-Sham) DFT. The off-shot is that one reduces the ground-state problem, which is a variational problem over the high-dimensional space of $N$-particles wave-function, to a system of $N$ (coupled) non-linear eigenvalue problems on the single-particle space. Thus, provided that the exchange-correlation potential can be efficiently evaluated (or approximated), this approach significantly reduces the dimension of the ground-state problem, thereby placing accurate solutions within the reach of current technologies.


\section*{Data availability}
No datasets were generated or analysed during the current study.

\section*{Competing interests}

The authors has no competing interests to declare that are relevant to the content of this article.


\bigskip
\end{document}